\documentclass{aa}
\usepackage{graphicx}
\usepackage{txfonts}
\usepackage{natbib}
\usepackage{verbatim}
\usepackage{txfonts,epsfig,graphicx,url,twoopt}
\usepackage[usenames,dvipsnames]{color}
\usepackage[breaklinks=true]{hyperref}
\hypersetup{colorlinks=true,citecolor=blue}
\bibpunct{(}{)}{;}{a}{}{,} 
\usepackage{hyperref}
\usepackage{color} 

\begin{document}
   \title{Planet-vortex interaction:\\How a vortex can shepherd a planetary embryo}

   \author{S.~Ataiee\inst{1,2,3}
          \and
          C.P.~Dullemond\inst{1}
          \and
          W.~Kley\inst{4}
          \and
          Zs.~Reg\'{a}ly\inst{5}
          \and
          H.~Meheut\inst{6}
          }

   \institute{Heidelberg University,
              Center for Astronomy,
              Institute for Theoretical Astrophysics,
              Albert Ueberle Str.~2, 69120 Heidelberg, Germany \email{sareh.ataiee@gmail.com}
         \and
            	School of Astronomy,
         	 Institute for Research in Fundamental Sciences (IPM),Tehran, Iran
         \and
    			 Department of Physics,
             Faculty of Sciences,
             Ferdowsi University of Mashhad, Mashhad, Iran
         \and
             Institut f\"{u}r Astronomie \& Astrophysik, Universit\"{a}t T\"{u}bingen,
             Auf der Morgenstelle 10, 72076, T\"{u}bingen, Germany
         \and
         	 Konkoly Observatory, 
         	 Research Center for Astronomy and Earth Science,
         	 Hungarian Academy of Sciences, Hungary
         \and
         	CEA, Irfu, SAp, Centre de Saclay, 
         	F-91191 Gif-sur-Yvette, France
             }

   \date{submitted Sep.20th 2013}

 \abstract{
 Anticyclonic vortices are considered as a favorable places for trapping dust and forming planetary embryos. On the other hand, they are massive blobs that can interact gravitationally with the planets in the disc.}
 {
 We aim to study how a vortex interacts gravitationally with a planet which migrates toward it or a planet which is created inside the vortex.}
 {
 We performed hydrodynamical simulations of a viscous locally isothermal disc using \texttt{GFARGO} and \texttt{FARGO-ADSG}. We set a stationary Gaussian pressure bump in the disc in a way that a large vortex is formed and maintained due to Rossby wave instability. After the vortex is established, we implanted a low mass planet ($[5,1,0.5]\times10^{-6}\mathrm{M_{\star}}$) in the outer disc or inside the vortex and allowed it to migrate. We also examined the effect of vortex strength on the planet migration by doubling the height of the bump and checked the validity of the final result in the presence of self-gravity.}
 {
 We noticed regardless of the planet's initial position, the planet is finally locked to the RWI created vortex in a 1:1 resonance or its migration is stopped in a farther orbital distance in case of a stronger vortex. For the model with the weaker vortex (our standard model), we studied the effect of different parameters such as background viscosity, background surface density, mass of the planet and different planet positions. In these models, while the trapping time and locking angle of the planet vary for different parameters, the main result, which is the planet-vortex locking, remains valid. We discovered that even a planet with a mass less than $5\times10^{-7}\mathrm{M_{\star}}$ comes out from the vortex and is locked to it at the same orbital distance. For a stronger vortex, both in non-self-gravitated and self-gravitating models, the planet migration is stopped far away from the radial position of the vortex. This effect can make the vortices a suitable place for continual planet formation under the condition that they save their shape during the planetary growth.
 }{}


   \maketitle
\section{Introduction}
Theoretical models suggest that vortices can develop in protoplanetary discs and help solving time scale problems in
the early planetary formation process and fast inward migration. The work by \cite{Li2000} gives the needed condition for Rossby wave instability and vortex formation in protoplanetary discs. According to this work, a surface density change over a radial distance of the order of the scale height of the disc is able to excite Rossby waves and eventually produce vortices. This condition can be provided by the disc spontaneously -- for instance due to the presence of a dead-zone -- or by the help of a gap-opening planet.
\cite{Klahr2003} showed that a protoplanetary disc with a radial gradient in the entropy becomes unstable and creates turbulence inside the disc. Owing to the turbulence-made pressure bumps, long-lasting vortices are formed. Vortices can also be established at the dead part of dead-zone edges because of a pile-up of matter due to different accretion rates \citep[e.g.][]{Regaly2011,Lyra2012a}. A massive planet, which opens a gap, can likewise produce a sufficiently steep density gradient at the gap edges and therefore vortices would be developed on both sides of the gap \citep[e.g.][]{Li2005,DeVal2007,Ataiee2013,Fu2014}.

It has been shown that anticyclonic vortices trap dust particles and boost planet formation \citep{Barge1995,Klahr1997,Chavanis2000,DeLaFuenteMarcos2001,Johansen2004,Inaba2006}. The Coriolis force and the resultant higher pressure at the center of anticyclonic vortices help to bring and accumulate the dust particles inside the vortices \citep[e.g.][]{Zhu2014}. In a twin work by \cite{Lyra2009a,Lyra2009} the possibility of planet formation by vortices was studied. In the former work, they studied the planet formation at the high pressure regions of a disc holding a Jupiter-mass planet in the presence of self-gravity. The results show that super-Earths can be formed inside the vortices generated at the outer edge of the gap. In the latter work, they used a similar approach to study the formation and evolution of vortices created at the edge of a dead-zone and concluded that the dust accumulation inside vortices is so effective that objects of planetary mass can be formed in five orbits. Nevertheless, the effect of particle collision, erosion or fragmentation has not been included in the mentioned work.

Anticyclonic vortices behave as ''blobs`` of matter and interact with the planets in the disc. \cite{Koller2003} noticed that for small smoothing lengths in the potential of the planet, some vortices are formed out of the planet Roche lobe and propagated in the co-orbital region. They stated that the vortices exert significant torques on the planet. \cite{Li2009} and \cite{Yu2010} confirmed this statement by showing that the torques from the vortices created at the outer edge of a planet-carved gap in weakly viscous discs produce a slight outward migration of the planet.

Considering vortices, as either a convenient birth place for planetary embryos or a mechanism for decreasing planetary migration rate, introduces an important question: how does a planet embryo interact with a vortex? If the planet is formed inside a vortex, does it leave its birth place to allow further planetary formation in the vortex or does it stay until it becomes massive enough to open a gap? If a planet is made external to the vortex but during its migration meets the vortex, does it enter the vortex and disturb the planetary formation process? In this work, we aim to answer these questions by studying the migration of a low-mass planet in the presence of a stationary vortex created inside a Gaussian pressure bump. The order of the paper is as follows: In the next section (\ref{method}) we describe our model and the parameters used, in section \ref{result} we present our results and discuss them in Sec.~\ref{discuss}, and we finally conclude with Sec.~\ref{summary}.

\section{Method and setup}
\label{method}
Pressure bumps, which are long-term pressure enhancements, are known as suitable traps for both particles and planets. Various processes such as accumulation of gas at the dead-zone edges or ice line \citep{Kato2009}, zonal flows \citep{Johansen2009} or outer edge of a gap created by a massive planet, can produce pressure bumps. Due to different drag forces felt by the particles at both sides of a pressure bumps, dust can drift toward the pressure maxima and grow \citep{Haghighipour2003,Johansen2007,Johansen2011,Pinilla2012}. On the other hand, low-mass planets which do not disturb the surface density much, are able to find a position in a bump where the differential Lindblad and corotation torques balance and hence are trapped \citep{Lyra2009}. However, \cite{Regaly2013} showed that, for slightly massive planets, that can open a partial gap (at least $10\mathrm{M_{\oplus}}$) trapping in a pressure maximum only occurs in the presence of a large-scale vortex. A Rossby wave instability (RWI), which can be considered as the rotational equivalent of Kelvin-Helmholtz instability, may also be triggered in a pressure bump and produce anticyclonic vortices. \cite{Li2000} show that if 10\%-20\% radial density variation exists with a radial width not more than the thickness of the disc, RWI can be excited. The vortices created by RWI merge and form a single or a few large-scale vortices which are appropriate places to trap and grow particles. This process can lead to planetesimal formation \citep{Lyra2009,Lyra2009a,Regaly2011}.

The question that arises here is: What would happen to a planet getting close to a vortex-holding bump or a planet-embryo which is created inside a vortex? We intend to answer this question by performing hydrodynamical simulations of discs which contain an embedded planet and a large vortex. We are interested in a large single vortex because (a)~a large vortex is more probable to be observed than small ones \citep[an example of a possible observed vortex:][]{VanderMarel2013}, (b)~simulations show that RWI vortices are prone to merge into lower modes \citep[e.g.][]{Lyra2009,Meheut2012a} and (c)~analysing the torques between a planet and a single vortex is much easier than multiple vortices. We also study the effect of different parameters such as viscosity, planet mass and vortex strength.

\subsection{Code}
\label{code}
We performed our simulations with the GPU double-precision version of the \texttt{FARGO} code \citep{Masset2000} called \texttt{GFARGO} \footnote{http://fargo.in2p3.fr/spip.php?rubrique21}. This is a Zeus-based code \citep{Stone1992} that solves Navier-Stokes equations with full viscous tensor implemented and continuity equation in a cylindrical coordinate system for a two-dimensional Keplerian disc at the presence of a central object and embedded planets. The continuity equation and the equation of motion read

\begin{eqnarray}
	&&\frac{\partial \Sigma}{\partial t} + \nabla . (\Sigma \vec{v}) = 0 \label{conti}\\
	&&\frac{\partial \vec{v}}{\partial t}+(\vec{v} . \nabla)\vec{v} = -\frac{1}{\Sigma} \nabla P - \nabla \Phi + \nabla . \mathbf{T} \label{navie}
\end{eqnarray}

\noindent{where $\vec{v}$ is the velocity vector, $P$ is the pressure and $\Phi$ is the gravitational potential of the star and planet,
it contains the indirect terms caused by the non-inertial coordinate system. In the case of a self-gravitating disk the corresponding accelarations
are directly added to the right hand side of Eq.~\ref{navie} \citep{2008ApJ...678..483B}.
The components of $\mathbf{T}$, the viscous stress tensor, are given as}

\begin{eqnarray}
	&& T_{rr}= 2 \Sigma \nu \left[ \frac{\partial v_{r}}{\partial r} - \frac{1}{3} (\nabla . \vec{v}) \right]\\
	&& T_{\theta \theta} = 2 \Sigma \nu \left[ \frac{1}{r} \frac{\partial v_{\theta}}{\partial \theta} + \frac{v_{r}}{r} - \frac{1}{3} (\nabla . \vec{v}) \right] \\
	&& T_{r\theta} = T_{\theta r} = \Sigma \nu \left[ r \frac{\partial}{\partial r}(\frac{v_{\theta}}{r}) + \frac{1}{r} \frac{\partial v_{r}}{\partial \theta} \right].
\end{eqnarray}

The isothermal equation of state is applied with a radial temperature profile which is determined through the aspect ratio or an arbitrary radial sound speed profile. 

We modified the code to enable us  (a) to insert a planet into the simulation at our desired time and angular position, (b) to set up a steady Gaussian bump for the viscosity, and (c) to start the velocity perturbation at a specific time. In this code, the calculations are performed in a coordinate system that is centred on the star. In the model Massive Vortex (SG), where we tested the effect of self-gravity, we used \texttt{FARGO-ADSG} that is capable to handle disc self-gravity and adiabatic thermodynamics.

To assure our results are not code-dependent, we tested some of our models with the code \texttt{RH2D} \citep{Kley1989,Kley1999} independently. The results of the both codes were identical.

\subsection{Numerical setup for the standard model}
\label{Standard}
We considered a locally isothermal disc with constant background surface density $\Sigma_{BG}=5 \times 10^{-4}\ [M_{\star}/r_{0}^2]$ and kinematic viscosity $\nu_{BG}=10^{-5}\ [r_{0}^2 \Omega_{k}(r_{0})]$ where $\Omega_{k}$ is Keplerian angular velocity. To avoid singular behaviour in the gravitational potential of the planet, we used a smoothing parameter $\epsilon=0.6H(r_{p})$ as

\begin{eqnarray}
	\label{smooth}
	\phi_{p}=-\frac{GM_{p}}{(r^2+\epsilon ^{2})^{1/2}}
\end{eqnarray}

\noindent where $M_{p}=5 \times 10^{-6} M_{\star}$ is the mass of the planet that is equivalent to $1.7M_{Earth}$ in a system with a solar mass star.
We used a fixed non-rotating 2D polar coordinate $(r,\theta)$ with 256 logarithmic grids in radial and 512 equidistant grids in azimuthal direction. The disc was extended from $0.5r_{0}$ to $1.5r_{0}$ and the wave-damping boundary condition \citep{DeVal-Borro2006} was applied to both inner and outer boundaries. In the models in which we initially start with the planet inside the vortex, the planet was introduced after 300 orbits when a large vortex was formed. In the rest of the models, the planet was introduced to the disc at the beginning  but its potential was slowly switched on during the first 10 orbits. We continued the simulations until the planet passed the bump and reached the inner boundary or until it was trapped inside the bump for more than 1000 orbits.

\subsection{Making a vortex}
\label{Mvor}
\begin{figure}
	\begin{center}
		\includegraphics[width=8cm]{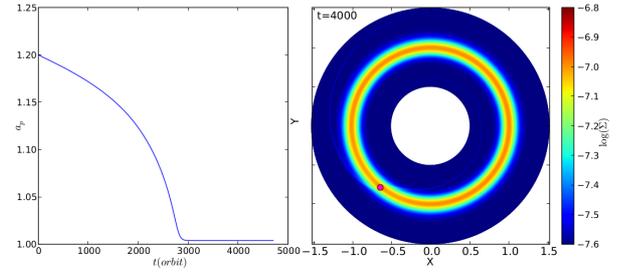}
		\caption{Semi-major axis evolution (left) and surface density at $t=4000$ orbits for \textit{Width06}. Because of a wide bump, no vortex is formed.}
		\label{width6}
	\end{center}
\end{figure}

In order to produce a single vortex and study its interaction with a planet, we need to have a controlled formation condition which enables us to understand the effect of each parameter. For instance, because of continuous gas accumulation at the edge of a dead-zone, the different parameters of the density bump, such as height and width, are changing in time and consequently influence the vortex \citep[e.g.][]{Regaly2013}. Therefore, we embedded a Gaussian density bump -- corresponding to a pressure bump in a locally isothermal disc -- into a smooth disc and altered the disc and planet properties while we used fixed values for the width and position of the bump. To prevent the bump smoothing out in time, we adjusted the viscosity profile to provide a physical mechanism in the system for the higher surface density in the bump. Specifically, we prescribed the following fixed profile for the viscosity
\begin{equation}
	\label{visc}
	\nu(r) = \frac{\nu_{BG} (r)}{1+a\ \exp(-\frac{(r-r_{0})^2}{2 c^2})} \,,
\end{equation}
\noindent where $\nu_{BG}(r)$ is the background viscosity of the disc and $a$ is the height of the bump that controls the amount of density/viscosity enhancement/reduction in the bump, $r_{0}$ represents the location of the bump in the disc and is used as the length scale in our simulations. Finally, $c$ controls the width of the bump. A dead-zone edge, where a viscosity reduction or enhancement happens, can be considered as a more realistic example of such density-viscosity adaptation
\citep{Lyra2009}. The disc was considered flat with the constant aspect ratio $h=0.05$. In the main models the bump parameters are fixed to $a=1$ and  $c=H(r_{0})=h(r_{0})r_{0}=0.05$ with $H(r)$ being the pressure scale height, except some test models in which we examine the effect of vortex strength on the planet migration. Our choice of $c=0.05$ is the marginal value for RWI formation based on the threshold presented by \cite{Li2000}. For a wider bump, we do not expect vortex formation. We tested this criterium in the test model~Width06 (see Table~\ref{table:1}) and the result in Fig~\ref{width6}~ shows no vortex even after 4000 orbits.

Considering the imposed viscosity profile we set up a Gaussian density profile, $\Sigma(r)$,  at the start of the simulations
\begin{equation}
	\label{dens}
	\Sigma(r) = \Sigma_{BG}(r) \left[1+a\ \exp(-\frac{(r-r_{0})^2}{2 c^2})\right] .
\end{equation}

We ignited the RWI with subsequent vortex formation either by adding an initial perturbation to the radial velocity inside the bump or by the planet itself. In the models without initial planet inside the disc, we added an initial random perturbation

\begin{equation}
	\label{velper}
	v_{pert}=10^{-4}rand(1)\exp\left(-\frac{(r-r_{0})^2}{2c^2}\right)
\end{equation}

\noindent to the radial velocity\footnote{The unit of the velocity perturbation is $[r_{0}/(2\pi /\Omega_{K}(r_0))]$ which is the code unit for velocity.}. $rand(1)$ is a function that produces a random number between 0 and 1. Our choice of random initial perturbation prevents the excitation of a specific mode. Because our specific set-up for the bump, the vortex neither gets destroyed nor migrates. In the cases with the planet initially inside the disc but out of the bump, the perturbation, which excites the RWI and triggers subsequent vortex formation, is created by the planet. 

\begin{figure}
	\begin{center}
		\includegraphics[width=9cm]{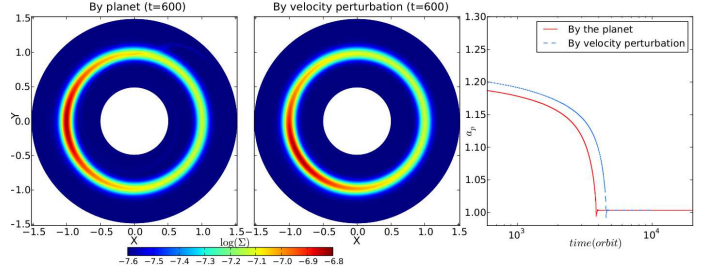}
		\caption{Comparison between two models, one where the vortex formation is ignited by the planet (left) and one where it is excited by the radial velocity perturbation (middle). The left and middle panels show the surface density after 600 orbits when a single vortex has been established. In the right panel, $a_{p}$ is the variation of planet semi-major axis.}
		\label{pert}
	\end{center}
\end{figure}

Although these two different methods might result in different initial azimuthal position and shape of the single vortex, the final trapping position of the planet relative to the vortex was not affected by the initialization method. A typical outcome of our simulations (here the standard model) is shown in Fig.~\ref{pert}, where we compare the density distribution of two runs with and without initial velocity perturbation (after eq.~\ref{velper}), and the migration history of an embedded planet of mass $M_{p}=5 \times 10^{-6} M_{\star}$. The planet was initialized at a distance $r_{p}=1.2 r_0$. 

\subsection{Resolution}
\label{resol}
\begin{figure}
	\begin{center}
		\includegraphics[width=9cm]{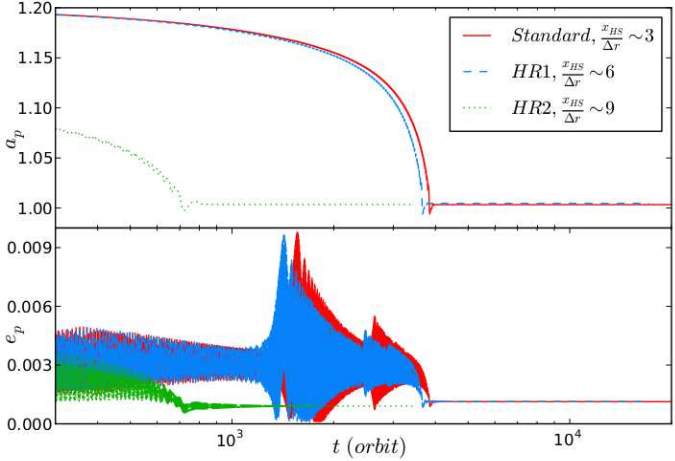}
		\caption{The effect of numerical resolution on the planet migration (upper panel) and planet eccentricity (lower panel). The trapping position of the planet is not sensitive to the resolution while the trapping time can vary. In HR2, because the outer boundary is at 1.3 (table \ref{table:1}), we set the planet slightly inner than the other two models to avoid the numerical issues.}
		\label{res}
	\end{center}
\end{figure}

To optimize the resolution considering run-time and numerical convergence, we repeated our standard model ($n_{r}\times n_{s}=256\times 512$) with two higher resolution models (HR1 and HR2 models in table \ref{table:1}). We compared the half-horseshoe width $x_{HS}$ with the cell width at the bump using the relation by \cite{Paardekooper2010} for low-mass planets. The half-horseshoe width was resolved by $\sim 3$, $\sim 6$ and $\sim 9$ cells in the standard, HR1 and HR2 models respectively. Figure \ref{res} shows that while the resolution can affect the migration rate and the trapping time, it does not change the trapping position and long-term behavior of the planet. \cite{Paardekooper2010} showed that the models with half-horseshoe width resolved through 6 cells are in good agreement with the analytical calculations. Moreover, Fig.\ref{res} presents identical trapping positions for our standard and higher resolution models. Thus we assure that the corotation torque is resolved for all numerical resolution. Therefore, we applied the resolution of our standard model to the rest of the simulations. 

\subsection{Parameters we study}
We studied how trapping of the planet is changed due to variation of some physical parameters such as: background viscosity, background surface density, planet mass, different initial position of the planet. We altered the background viscosity between $ 10^{-5},~10^{-6}$ and $10^{-7}$ which corresponds to $\alpha=4\times10^{-3},~4\times10^{-4}$ and $4\times10^{-5}$ in $\alpha$-prescription of viscosity \citep{Shakura1973}. We chose a lower $\Sigma_{BG}=10^{-4}$ and a higher $10^{-3}$ background surface density than the standard model. Because the vortex can be a suitable place for producing planet embryos, we tried lower planet masses $M_{p}=1\times 10^{-6}$ and $5\times10^{-7}$ (equivalent to $0.3$ and $0.17$ Earth mass in a solar-like system ) with their initial position inside the vortex to have an estimation of embryos mass that can stay inside a vortex. Because the value of horseshoe drag is affected by the resolution, we adjusted the radial span to always keep the resolution ($x_{HS}/\Delta r\sim3$) equal to the standard model. To check if the final trapping position of the planet depends on the initial position of the planet, we repeated the standard model with the planet in different positions toward the vortex. Table \ref{table:1} summarizes the important parameters of all models. 

\begin{table*}
\caption{Models with the parameters we studied. Test models are ones we used in section \ref{discuss} to clarify the results. $r_{in}$ and $r_{out}$ are inner and outer boundaries.}
\label{table:1}
\centering
\begin{tabular}{c c c c c c c c c c c}
\hline\hline
Model & $\nu_{BG}[r_{0}^2 \Omega_{k}(r_{0})]$  & $\Sigma_{BG}[M_{\star}/r_{0}^2]$ & $M_{p}[M_{\star}]$ & $(\frac{r_{init}}{r_{0}},\theta_{init}^{\circ})_{p}$ & $N_{r} \times N_{s}$ & $(r_{in},r_{out})[r_{0}]$ & $a$ & $c[r_{0}]$ & $M_{disc}[M_{\star}]$\\
\hline
Standard & $10^{-5}$ & $5 \times 10^{-4}$ & $5 \times 10^{-6}$ & $(1.2,0)$ & $256 \times 512$ & (0.5,1.5) & 1 & 0.05 & 0.0070\\
Standard (PiV)\tablefootmark{*}& $10^{-5}$ & $5 \times 10^{-4}$ & $5 \times 10^{-6}$ & $(1,80)$ & $256 \times 512$ & (0.5,1.5) & 1 & 0.05 & 0.0070\\
Visc1 & $10^{-6}$ & $5 \times 10^{-4}$ & $5 \times 10^{-6}$ & $(1.2,0)$ & $256 \times 512$ & (0.5,1.5) & 1 & 0.05& 0.0070\\
Visc2 & $10^{-7}$ & $5 \times 10^{-4}$ & $5 \times 10^{-6}$ & $(1.2,0)$ & $256 \times 512$ & (0.5,1.5) & 1 & 0.05 & 0.0070\\
Dens1 & $10^{-5}$ & $ 10^{-3}$ & $5 \times 10^{-6}$ & $(1.2,0)$ & $256 \times 512$ & (0.5,1.5) & 1 & 0.05& 0.014\\
Dens2 & $10^{-5}$ & $ 10^{-4}$ & $5 \times 10^{-6}$ & $(1.2,0)$ & $256 \times 512$ & (0.5,1.5) & 1 & 0.05 & 0.0014\\
Mass1 & $10^{-5}$ & $5 \times 10^{-4}$ & $ 10^{-6}$ & $(0.99,359)$ & $512 \times 1024$ & (0.6,1.4) & 1 & 0.05& 0.0070\\
Mass2 & $10^{-5}$ & $5 \times 10^{-4}$ & $ 5 \times 10^{-7}$ & $(0.99,69)$ & $512 \times 1024$ & (0.7,1.3) & 1 & 0.05& 0.0070\\
\hline
\hline
&&&&Test models\\
\hline
HR1 & $10^{-5}$ & $5 \times 10^{-4}$ & $5 \times 10^{-6}$ & $(1.2,0)$ & $576 \times 1152$ & (0.5,1.5) & 1 & 0.05& 0.0070\\
HR2 & $10^{-5}$ & $5 \times 10^{-4}$ & $5 \times 10^{-6}$ & $(1.1,0)$ & $576 \times 1152$ & (0.7,1.3) & 1 & 0.05& 0.0070\\
$O_{n}$\tablefootmark{**} & $10^{-5}$ & $5 \times 10^{-4}$ & $5 \times 10^{-6}$ & $(1.05,80+n)$ & $256 \times 512$ & (0.5,1.5) & 1 & 0.05& 0.0070\\
$I_{n}$ \tablefootmark{**}& $10^{-5}$ & $5 \times 10^{-4}$ & $5 \times 10^{-6}$ & $(0.95,80+n)$ & $256 \times 512$ & (0.5,1.5) & 1 & 0.05& 0.0070\\
Height2 & $10^{-5}$ & $5 \times 10^{-4}$ & $5 \times 10^{-6}$ & $(1.2,0)$ & $256 \times 512$ & (0.5,1.5) & 2 & 0.05 & 0.0078\\
Height2 (SG)\tablefootmark{*} & $10^{-5}$ & $5 \times 10^{-4}$ & $ 5 \times 10^{-6}$ & $(1.2,0)$ & $256 \times 512$ & (0.5,1.5) & 2 & 0.05& 0.0078\\
Visc1 (FP)\tablefootmark{*} & $10^{-6}$ & $5 \times 10^{-4}$ & $5 \times 10^{-6}$ & $(1.001534,0)$ & $256 \times 512$ & (0.5,1.5) & 1 & 0.05 &0.0070\\
Width06 & $10^{-5}$ & $5 \times 10^{-4}$ & $5 \times 10^{-6}$ & $(1.2,0)$ & $256 \times 512$ & (0.5,1.5) & 1 & 0.06 & 0.0072\\
\hline
\end{tabular}
\tablefoot{
\tablefoottext{*}{PiV, FP and SG stand for \textit{Planet inside Vortex}, \textit{Fixed Planet} and \textit{Self-Gravity}. Height2 (SG) model is run with the self-gravitating version of \texttt{FARGO} with the initial condition exactly identical to \textit{Height2} model except considering disc self-gravity.\\}
\tablefoottext{**}{$n$ is the azimuth of the planet toward the vortex in degree.}
}
\end{table*}

\section{Results}
\label{result}
\subsection{Standard model}
The red line in Fig.\ref{res} shows the evolution of the semi-major axis of the planet in our standard model. The planet is started out of the bump at $r_{init}=1.2r_{0}$ and migrates toward the bump. Due to the steeper surface density slope in the outer part of the bump, the planet migrates faster when it approaches the vortex, and after a severe interaction with the vortex at $\sim 3000$ orbits it is trapped by the vortex. Not only is the planet trapped by the vortex, but it is also locked to it by keeping a constant distance from the vortex center ($\approx$90 degrees from the vortex center). One might find this result in contradiction with \cite{Regaly2013} in which they showed a temporary trapping of the planet. In their work, they studied a $10~M_{\oplus}$ planet which opened a partial gap and therefore destroyed the vortex as it came close to it. Because our study deals with low-mass planets that do not open a gap and also a vortex that is long-lived, the planets are trapped permanently.

\begin{figure*}
	\begin{center}
		\includegraphics[width=18cm]{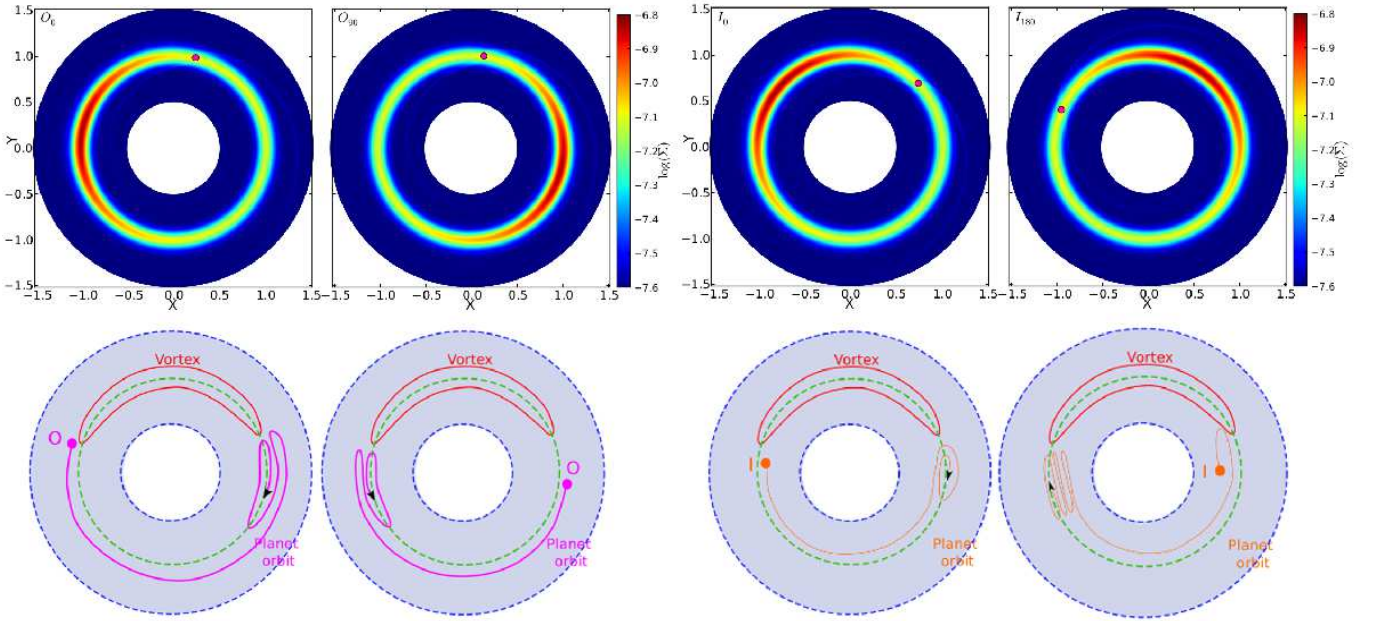}
		\caption{The upper panels demonstrate the final trapping positions of two \textit{O-models} (left panels) and two \textit{I-models} (right panels). In the lower illustrations, we draw the path of the planet from the times marked in Fig.~\ref{semiInOut} with ''O`` and ''I`` until they are trapped. For clarity, we exaggerated in showing the migration of the planets.}
		\label{InOut}
	\end{center}
\end{figure*}

\begin{figure}
	\begin{center}
		\includegraphics[width=9cm]{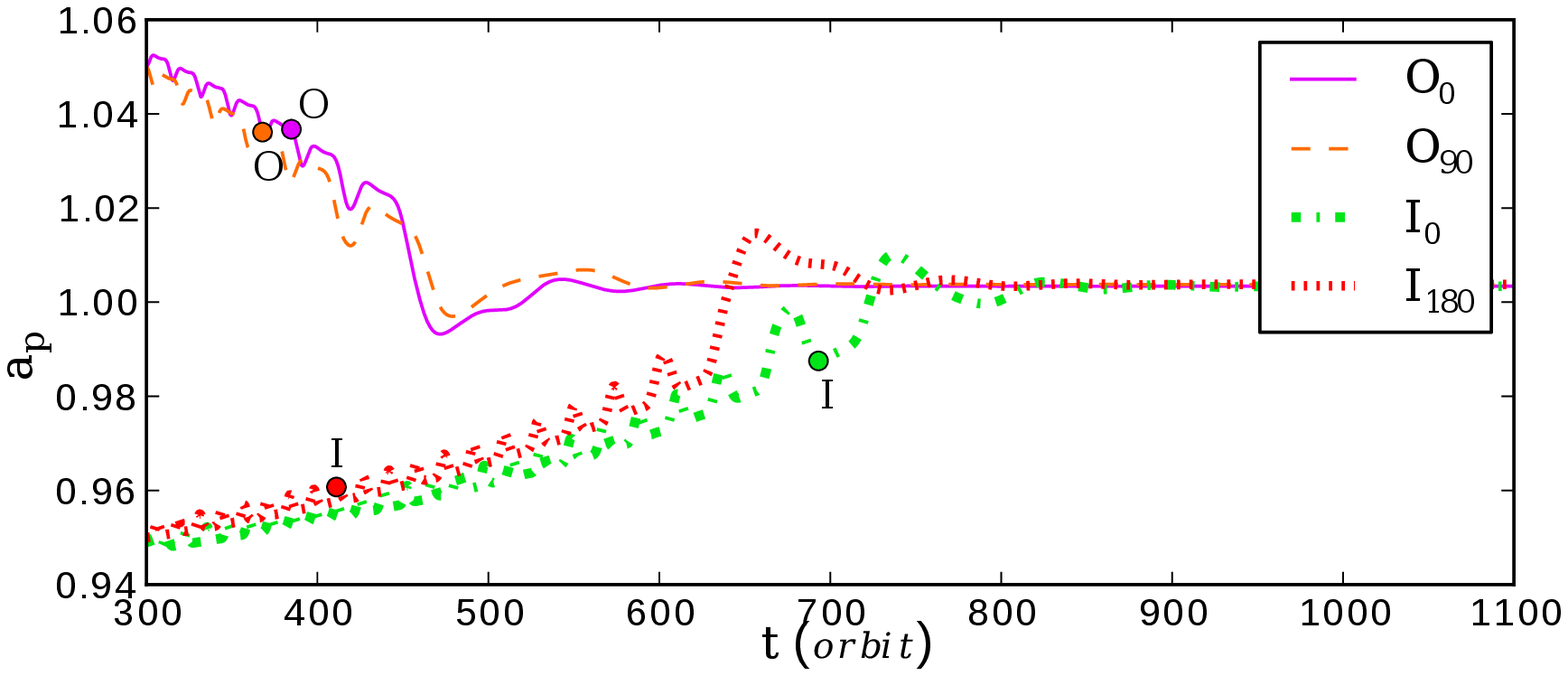}
		\caption{Semi-major axis evolution of four selected models displayed in Fig.~\ref{InOut}.}
		\label{semiInOut}
	\end{center}
\end{figure}

Checking the dependency of the results on the initial planet position, we examined sixteen test models with the planets implanted slightly outward or inward of the vortex radial position and different azimuth toward the vortex center . In these simulations, we saved the outputs every $1/10\mathrm{th}$ of an orbit to follow the planet path precisely. We named the models with planet initialized at $a_{i}=1.05r_{0}$, $O_{n}-\mathrm{models}$ and the ones with planet at $a_{i}=0.95r_{0}$, $I_{n}-\mathrm{models}$ where $n$ represents the planet-vortex azimuthal difference. We found that while the planet is radially trapped in the similar position to the standard model, the locking side can be either trailing or leading. As seen in table~\ref{table:3}, no relation can be found between the initial position of the planet and the final locking side. In Fig.~\ref{InOut}, we display the surface density of four models and below of each panel, we draw the corresponding path of the planet from the times marked in Fig.~\ref{semiInOut}. The final locking position seems to be given by the relative position of the planets upon close approach that cannot be predicted a priori.

\begin{table}
\caption{The final azimuthal position of the planet in O and I models. L/T stand for leading and trailing.}
\label{table:3}
\centering
\begin{tabular}{c c c c c c c c c}
\hline\hline
Model & 0 & 45 & 90 & 135 & 180 & 225 & 270 & 315\\
\hline
O-model & T & T & L & L & T & T & T & L\\
I-model & T & T & L & T & L & L & T & L\\
\hline
\end{tabular}
\end{table}

\begin{figure}
	\begin{center}
		\includegraphics[width=9cm]{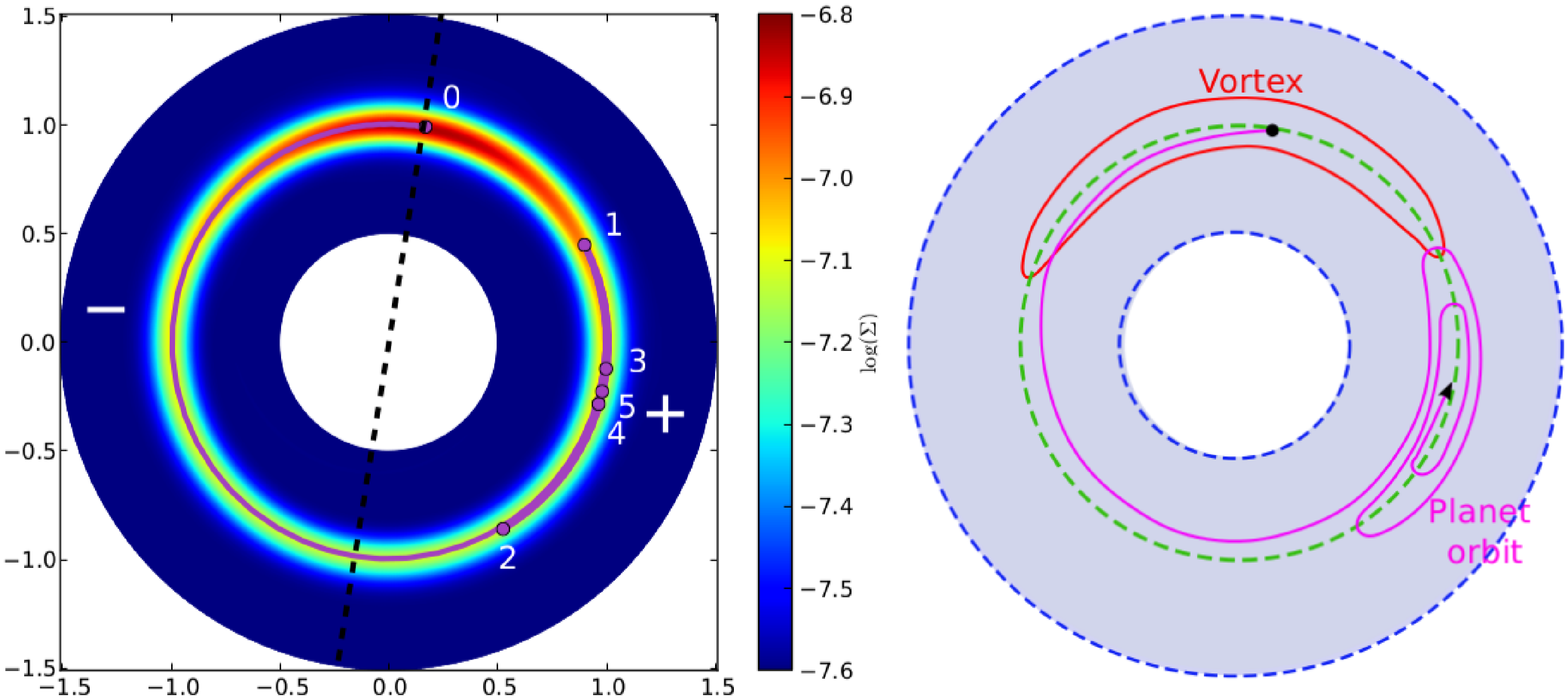}
		\includegraphics[width=9cm]{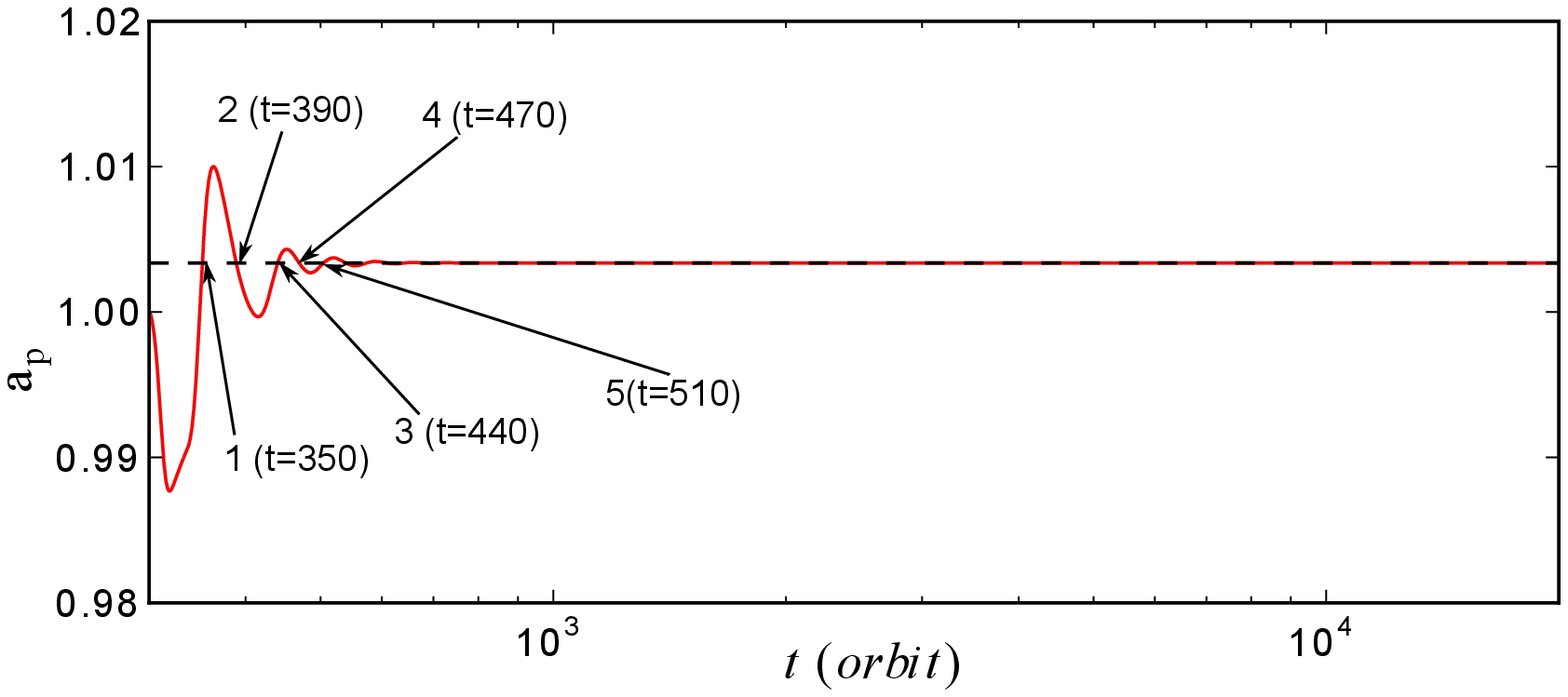}
		\caption{Planet trapping  in the frame co-rotating with the vortex. The left upper panel demonstrates a planet initially set inside the vortex (position 0), comes out and moves to the position 1 and then return back to 2. Afterward, it oscillates around position 5 until it is locked to the vortex at this location. We illustrate the planet's behaviour in the right upper panel in a more exaggerated way to show the radial migration of the planet as well. The lower panel shows the semi-major axis of the planet by time. The numbers denote the times of the marked positions in the left upper panel. We marked the positive and negative torques from the vortex by the $+$ and $-$ signs. The indirect torque from the star has the opposite sign of the vortex torque (Sec.\ref{discuss}). }
		\label{trace}
	\end{center}
\end{figure}

To see what happens for a planet which is born and becomes massive inside a vortex, we started the simulation by implanting the planet inside the vortex. In the upper panel of Fig.\ref{trace} we track the planet's motion in a frame co-rotating with the vortex. The planet gradually leaves the vortex and never enters back to it while oscillating around a specific location where it is trapped eventually. We plotted the semi-major axis of the planet in the lower panel of Fig.\ref{trace}. It shows that the planet during the first $~100$ orbits migrates inward and then outward. After that, the planet oscillates radially around a position where it stayed there until the end of the simulation. We sketch the orbital evolution of the planet in the illustration in Fig.\ref{trace}.

\subsection{Parameter study}
\label{param}
\begin{figure}
	\begin{center}
		\includegraphics[width=9cm]{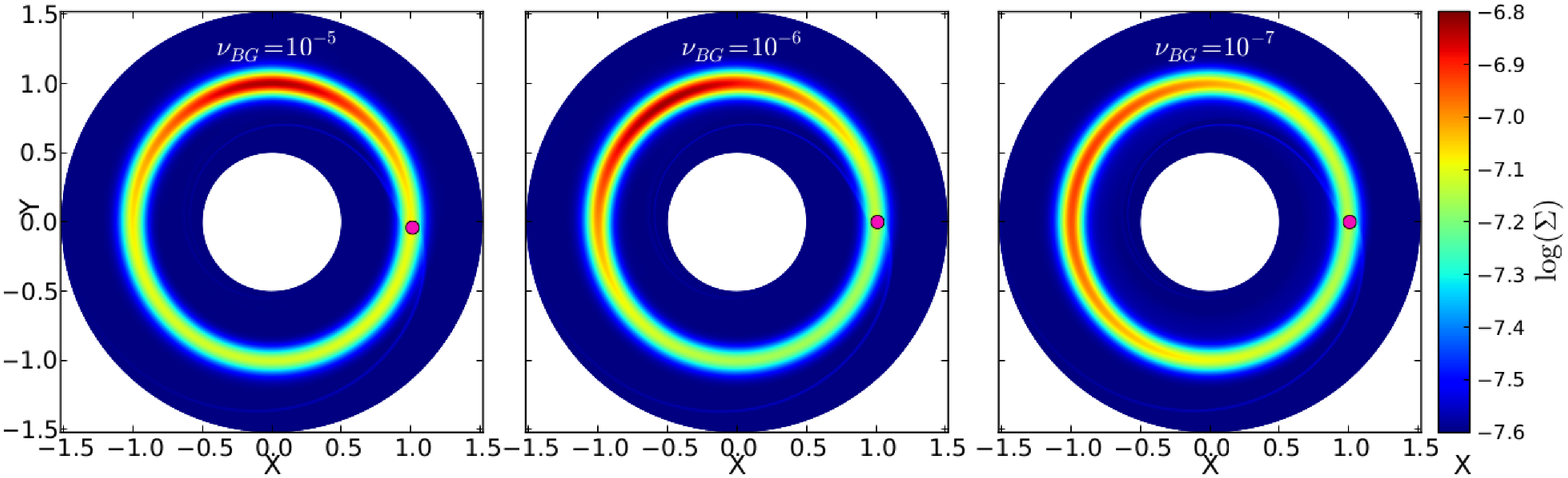}		
		\includegraphics[width=9cm]{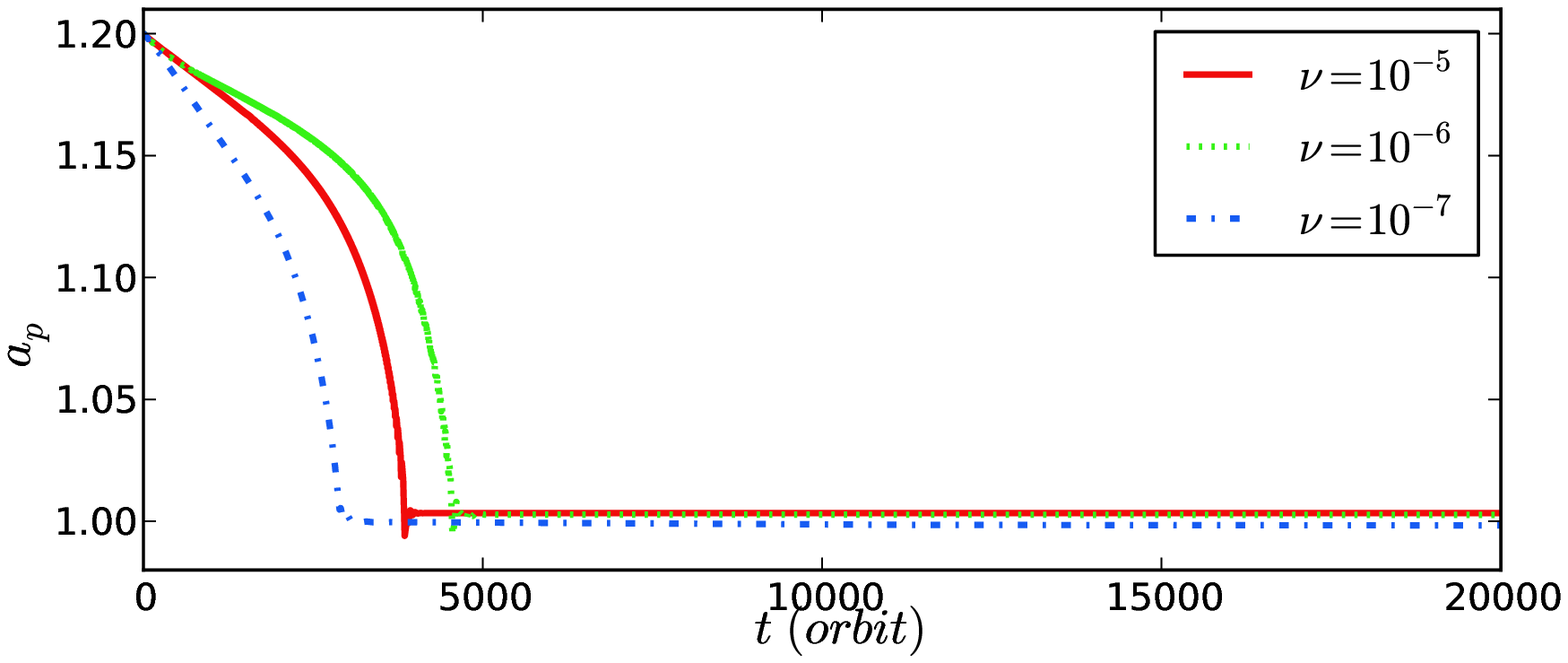}
		\caption{The trapping of the planet in discs with different background viscosity. The upper panels demonstrate the locking position of the planet toward the center of the vortex. The lower panel shows the change of planet semi-major against time. }
		\label{semvis}
	\end{center}
\end{figure}

Figure~\ref{semvis} shows the results for discs with different background viscosity. While the semi-major axis of the trapped planet is nearly the same for all cases, the azimuthal distance between the planet and the vortex center differs when altering the viscosity. This can be explained by considering the shape of the vortices. As the viscosity decreases, the vortex becomes more elongated. Therefore, the planet which is locked to the tail of the vortex is located farther from the vortex-center. The lower panel of Fig.~\ref{semvis} shows that the planet migration rate and trapping time also depend on the background viscosity. In all of our models with the planet starting outside of the bump, the planet's inward migration is accelerated as it approaches the bump. This happens because both of the Lindblad and vortensity-related component of corotation torques becomes more negative as the result of steeper surface density \citep[see][for more detail]{2012ARA&A..50..211K,Paardekooper2010}. Different migration rates is the consequence of corotation torque saturation \citep[see][as a good review]{2013LNP...861..201B}.

\begin{figure}
	\begin{center}
		\includegraphics[width=9cm]{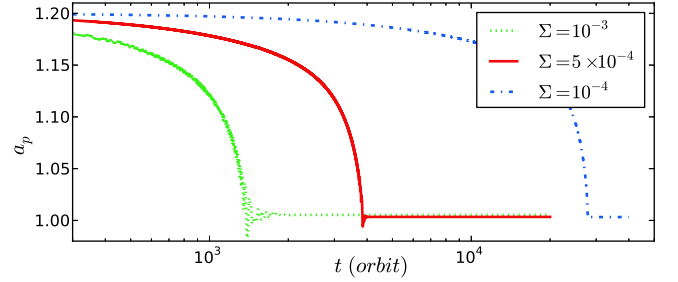}
		\caption{The evolution of planet semi-major axis for different background surface densities. The higher amplitude and longer oscillation of the planet semi-major axis before trapping shows the stronger interaction between the planet and the vortex in the case with $\Sigma=10^{-3}$.}
		\label{bgsurf}
	\end{center}
\end{figure}

The background surface density is another parameter we altered to see whether the mass of the vortex has a destabilizing effect. Figure~\ref{bgsurf} displays the evolution of planet semi-major axis in discs with different background surface densities. In the case with the highest value $\Sigma_{BG}=10^{-3}$, the migration is faster and the planet is trapped after a stronger interaction with the vortex compared to the standard model. For the lowest background surface density $\Sigma_{BG}=10^{-4}$, the planet migrates more slowly but is trapped eventually.

\begin{figure}
	\begin{center}
		\includegraphics[width=9cm]{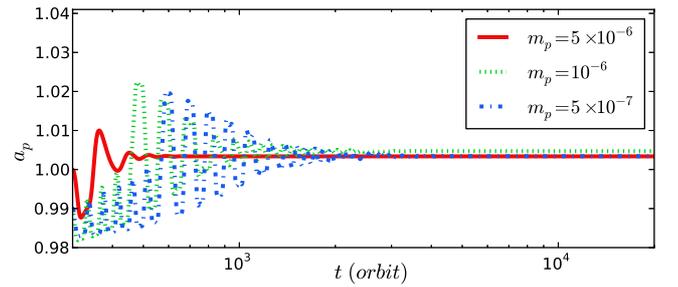}
		\caption{The evolution of planet semi-major axis for different planet masses.}
		\label{lowmass}
	\end{center}
\end{figure}

Figure~\ref{lowmass} shows that the three stages of migrating inward, outward and oscillating around the trapping position exist for lower planet masses as well. Although the oscillating time is longer for less massive planets, all of them get trapped finally.

\section{Discussion and torque analysis}
\label{discuss}

\subsection{Dominant torque components}
\label{dtcs}

\begin{figure}
	\begin{center}
		\includegraphics[width=7cm]{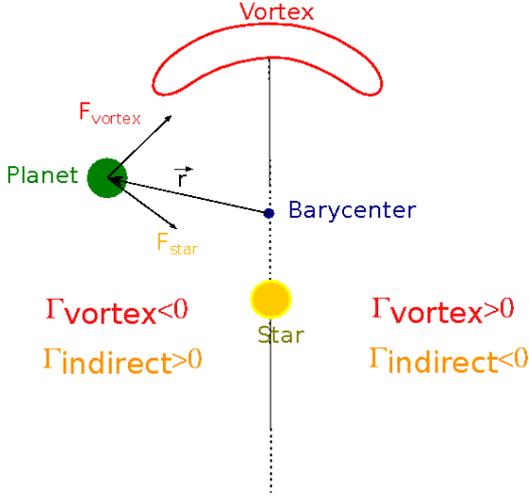}
		\caption{The star and the vortex torque on the planet in a coordinate centered on the barycenter. The distance between the barycenter and the star is much smaller than what is shown. One should keep in mind that  we illustrate the physics of our model in this drawing and it is not what the code uses for the calculations.}
		\label{SV}
	\end{center}
\end{figure}

To understand the reason of the planet locking to the vortex, either head or tail, we need to figure out which torques are exerted on the planet and which ones play the main role. Usually, in disc models with axisymmetric background, Lindblad and corotation torques \footnote{The corotation torque has two components: linear and non-linear (or horseshoe drag). \cite{Paardekooper2008} showed that the nonlinear component, if unsaturated, can dominate the migration of the planet. Therefore, when we use the term corotation torque in this work, we mean the horseshoe drag.} determine the migration of a planet. In the presence of a non-axisymmetric feature in the disc or a massive companion, the barycenter of the system is shifted away from the star and therefore, in the center of mass frame, the star exerts a torque on the planet due to its displacement from the barycenter. The schematic view of the star and the vortex torques are shown in Fig.~\ref{SV}. In the code, the calculation is done in the coordinate frame centred on the star though. The asymmetric feature in the disc accelerates the star by shifting the barycenter away and consequently, an indirect torque is exerted on the planet because of the accelerated coordinate frame. This indirect term in the accelerated frame is equivalent to the star torque in the center of mass frame and we will use them interchangeably for the convenience. We note that the vortex produces an azimuthal density gradient in the co-orbital region of the planet that might also modify the corotation contribution in the disc torque.

We choose the Visc1 model for torque analysis and generalise the discussion to the rest of the models. The Visc1 model, which is the standard model with $\nu=10^{-6}$, is an intermediate model considering the vortex length. In order to study the torques on the planet easier, we repeated the model with $\nu=10^{-6}$ in a corotating frame. The indirect torque is given by  

\begin{equation}
	\label{startorque}
	\Gamma_{\star}=-G x_{p} 	\sum_{i=0,j=0}^{Nr-1,Ns-1}\frac{\Sigma_{i,j}A_{i,j} y_{i,j}}{(x_{i,j}^2+y_{i,j}^{2})^{3/2}} + G y_{p} 	\sum_{i=0,j=0}^{Nr-1,Ns-1}\frac{\Sigma_{i,j}A_{i,j} x_{i,j}}{(x_{i,j}^2+y_{i,j}^{2})^{3/2}}
\end{equation}
\noindent where $A_{i,j}$ is the area of the cell $[i,j]$. In the corotating frame, the second term vanishes and $x_{p}=a_{p}$. The disc torque $\Gamma_{disc}$ is the gravitational torque from all disc cells on the planet in the frame centred on the star.

\begin{figure}
	\begin{center}
		\includegraphics[width=9cm]{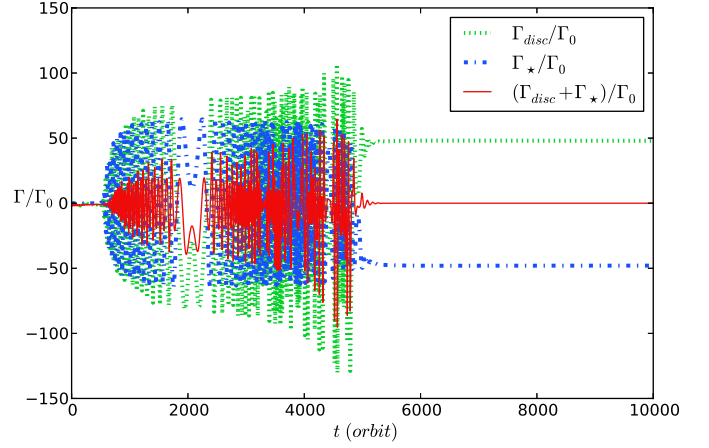}
		\caption{The normalized torques from the disc ($\Gamma_{disc}$) and the star ($\Gamma_{\star}$) on the planet in the model with $\nu=10^{-6}$. The solid line represents the net torque from both the disc and the star. One can see that the stellar and the disc torques are always opposite in sign.}
		\label{visc6co}
	\end{center}
\end{figure}

Figure \ref{visc6co} displays the star torque and the disc torque on the planet. The torques are normalized to $\Gamma_{0}$ which is defined as

\begin{equation}
\label{gam0}
\Gamma_{0}=\left(\frac{q}{h}\right)^{2}\Sigma(a_{init})a_{init}^{4}\Omega(a_{init})^{2}
\end{equation}

\noindent where $\Sigma(a_{init})$ and $\Omega(a_{init})$ are the surface density and Keplerian angular velocity at the initial location of the planet($a_{init}=1.2r_{0}$). $q$ represents the ratio of the planet mass to the star mass. The figure shows that when the planet is far from the vortex orbit, the disc exerts an oscillating torque on the planet due to the changing distance between the vortex and the planet. Because the system's barycenter is also rotating around the coordinate center, the star torque oscillates too. After about $5000$ orbits, the disc torque and the star torque balances and the planet is trapped.

While the balance of stellar and disc torques explains the radial location of the final trapping, the locking location in azimuth, here to the tail of the vortex, needs to be discussed. To answer this question, we need to find out how different torque components depend on the azimuthal position of the planet. The torques on the planet are derived from Lindblad resonances, corotation and horseshoe region, the vortex and the star displacement. 

\begin{figure}
	\begin{center}
		\includegraphics[width=9cm]{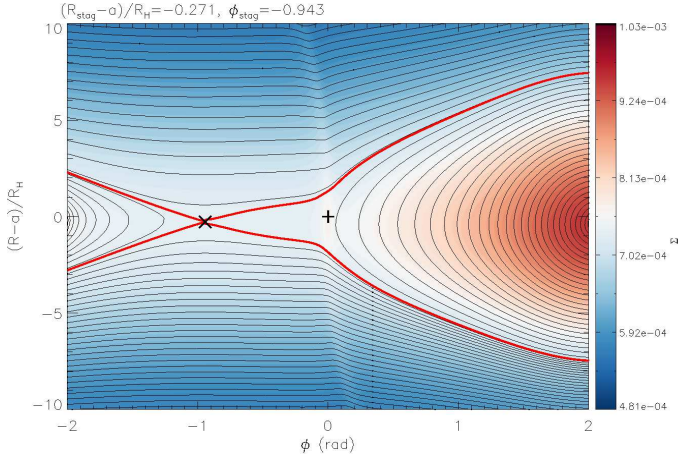}
		\caption{Surface density and streamlines for the Visc6 model for the trapped planet. The thick red line is the separatrix around the horseshoe region of the planet. The $+$ and $\times$ demonstrate the position of the planet and stagnation point respectively. The distances from the planet ($R-a$) are scaled by the Hill radius of the planet ($R_{H}$).}
		\label{visc6stream}
	\end{center}
\end{figure}

\begin{figure}
	\begin{center}
		\includegraphics[width=9cm]{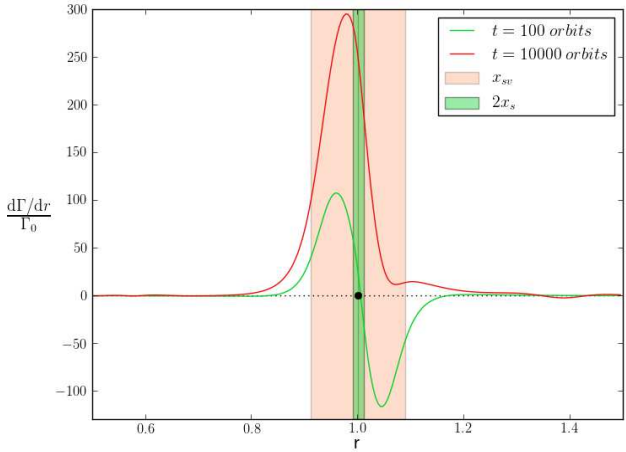}
		\caption{Radial torque distribution scaled by $\Gamma_{0}$ for the \textit{Visc1 (FP)} model. The green and red lines represent the torque distribution before the vortex was developed and and at the end of the simulation. The green area exhibits the analytically calculated horseshoe region for our planet in an axisymmetric disc. The orange area shows the larger horseshoe width obtained from Fig.\ref{visc6stream}. The position of the planet is also marked by a black dot. We note that we only plotted the torque from the disc and thus the integral of the red curve, which is non-zero, equals to $-\Gamma_{\star}$.}
		\label{tordis}
	\end{center}
\end{figure}

In Fig.\ref{visc6stream}, we show the area around the planet's orbit with more details including streamlines and horseshoe region. The spirals emanating from the planet, starting above and below the + symbol in Fig~\ref{visc6co}, give rise to the of Lindblad torques. The Lindblad torque arises from the location of the inner and outer Lindblad resonances which can be shifted by different parameters such as surface density or deviation of the velocity profile from Keplerian. On the other hand, the vortex changes the velocity profile of the neighbouring gas and pushes the open streamlines farther away in comparison with the no-vortex case (see deformation of streamlines outside of the red line in Fig.\ref{visc6stream}). According to the definition of the Lindblad resonances (where the gas angular frequencies in the planet's co-moving frame match the epicyclic frequencies), they cannot exist inside the vortex or horseshoe region where the streamlines are closed. To estimate the Lindblad torque after planet locking, we repeated the model with $\nu=10^{-6}$ in the corotating frame with a fixed planet located at the trapping position. Figure~\ref{tordis} shows the radial torque distribution for $t=100$ orbits and at the end of the simulation. The green area, which is denoted by $2x_{s}$, shows the analytically calculated horseshoe width and can be considered as the width of the horseshoe region before the vortex formation. The area colored by orange represents the larger horseshoe width at the end of the simulation. The Lindblad resonances contribution comes from the sum of torque values outside of the horseshoe region. If we consider the orange area as the horseshoe width and integrate the red curve in Fig.~\ref{tordis} over the radii outside of the horseshoe region, it gives us an estimate for the Lindblad torque. This value is $\Gamma_{L}/\Gamma_{0}\simeq 5$ which is $\sim 15\%$ of the total torque and shows that while the Lindblad torque is considerable, the main component of the torque belongs to the torque from the horse-shoe region.

The presence of the vortex both deforms the horseshoe region (red line in Fig.\ref{visc6stream}) and exerts an extra torque on the planet. The horseshoe width is much larger on the vortex side than the width on the other side. Not only does the asymmetric horseshoe region cause a large corotation torque on the planet, but the asymmetric azimuthal density distribution also produces a torque with the same sign as the corotation torque (positive in this case). Hence, the presence of the vortex inside the horseshoe region changes the torques such that the planet can be captured by the vortex. The vortex torque and corotation torque cannot be calculated separately because they are tightly related. For estimating the contributions of the vortex and corotation torques, we subtract the density at $t=100$ orbits --before the vortex formation-- from the surface density at the end of the simulation in order to obtain the density enhancement due to the vortex. Then, we calculate the vortex gravitational torque on the planet in the corotating frame using the below relation

\begin{equation}
	\label{gamVort}
	\Gamma_{V} = x_{p} \sum_{i,j=is,js}^{if, jf}	\frac{\Sigma_{ij}A_{ij} (y_{ij}-y_{p})}{((x_{ij}-x_{p})^2+(y_{ij}-y_{p})^2+(0.6h)^2)^{3/2}}
\end{equation}

\noindent where $[is,if]$ and $[js,jf]$ represent the cell indices of radial and azimuthal extension of the vortex. We estimate the vortex radial and azimuthal extension based on the density reduction as 1.7 of the background value. We note that this value is chosen in a way that could describe the shape of the vortex well enough. Because we were only interested on and estimation of the vortex mass, this simple method is fitted our aim. We avoid the singularity around the planet by adding the smoothing parameter $0.6h$ to the calculation. Using this method, the vortex torque contribution equals to $\Gamma_{V}/\Gamma_{f} \simeq 18\%$. The rest of the final torque which is $\simeq 67\%$ originates from the corotation torque.

\begin{figure}
	\begin{center}
		\includegraphics[width=9cm]{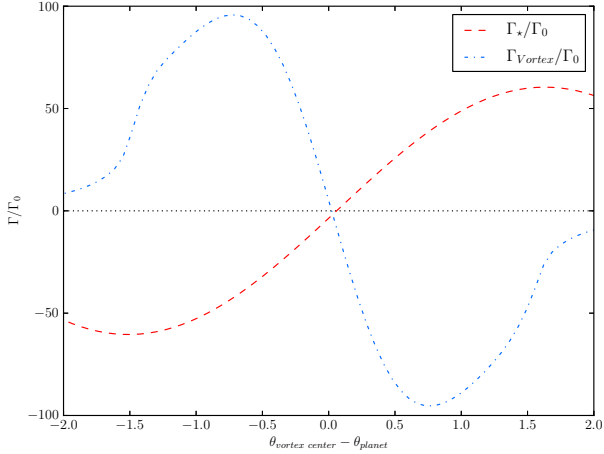}
		\caption{Star torque and vortex torque as a function of azimuthal distance between the planet and the vortex center for the Visc1 model.}
		\label{SVtorq}
	\end{center}
\end{figure}

The star torque is also a function of planet azimuth. It can be easily understood in the frame centred on the barycentre (see Fig.~\ref{SV}). As the planet rotates around the star, the angle between the star force $F_{\star}$ and planet position vector $\overrightarrow{r}$ changes. Because the torque depends on the sine of this angle, star torque has its maximum value where the planet azimuth toward the vortex is $\pi$. In Fig.~\ref{SVtorq}, we calculated and plotted the star torque and the vortex torque as a function of planet azimuth. As the results show, the distance between azimuthal locking position of the planet and the vortex is not necessarily $\pi$ but it depends on the vortex extend.

As the results of O and I models show, both end of the vortex wings are stable points. In the leading side, the vortex and corotation torques are negative and the star torque is positive. In contrast, in the trailing side, the vortex and corotation torques are positive and start torque is negative. How the planet chooses leading or trailing side is unclear to us but we think that the planet is trapped where it succeeds to adjust its horseshoe orbit to the vortex streamlines. Because the locking side of the planet has no observational consequence or influence on planet formation, we leave the investigation of this point to future study.

Similar torque analysis can be applied to the other models too. For example, in the model with a planet started at the vortex center, the planet initially migrates due to the Lindblad torques arisen from the Lindblad resonances out of the vortex. We note that while the corotation torque and the torque from the vortex can not exist inside the vortex (because the vortex is dominated to the planet horseshoe streamlines), the condition for the Lindblad resonances can be fulfilled outside of the vortex, where the gas elements do not have closed orbits. Because the vortex center is not a stable point, the planet comes out of the vortex (refer to Appendix~\ref{app1} for supplementary discussion). The migration of the planet continues until it reaches the point where the planet, star and vortex center are located on a straight line. Similar to the vortex center, this point is not an equilibrium position owing to the Lindblad torques. After that, the battle between the inward and outward migration continues until the planet finds the right place where the corotation, vortex and Lindblad torque can cancel out the torque from star.

\subsection{Strength of the vortex}
\label{massVort}
\begin{figure}
	\begin{center}
		\includegraphics[width=9cm]{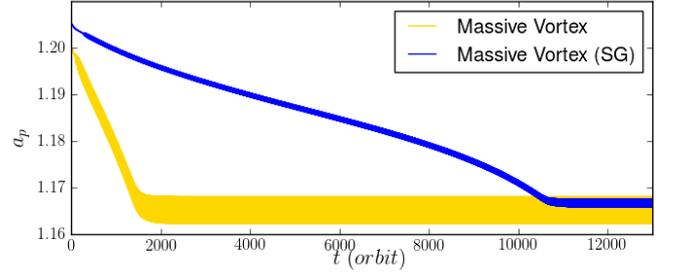}
		\caption{Semi-major axis evolution for the standard models but with double height for the bump (Height2 models). The yellow and blue lines represent the model with and without self-gravity, respectively.}
		\label{h2semi}
	\end{center}
\end{figure}

\begin{figure}
	\begin{center}
		\includegraphics[width=9cm]{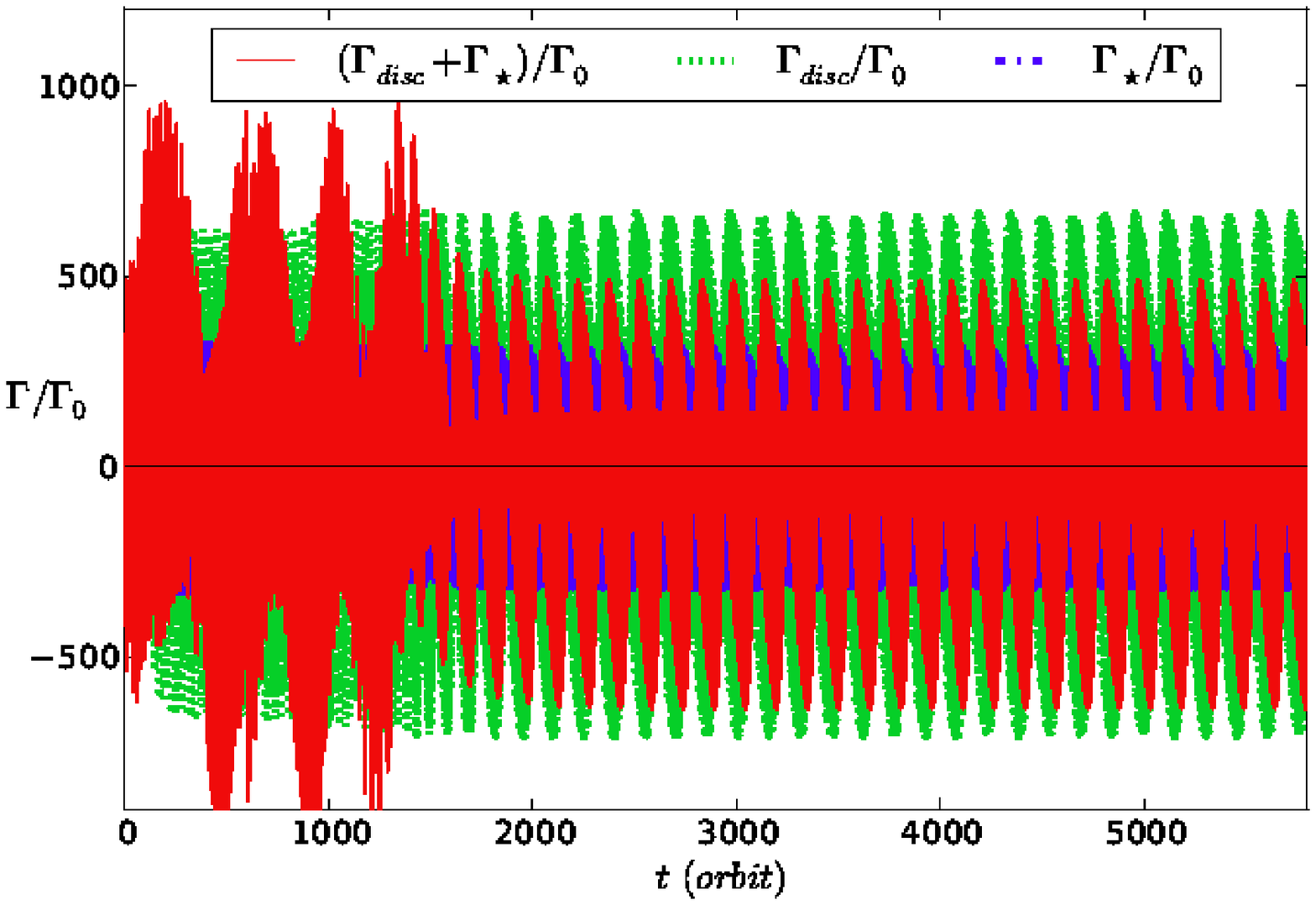}
		\includegraphics[width=9cm]{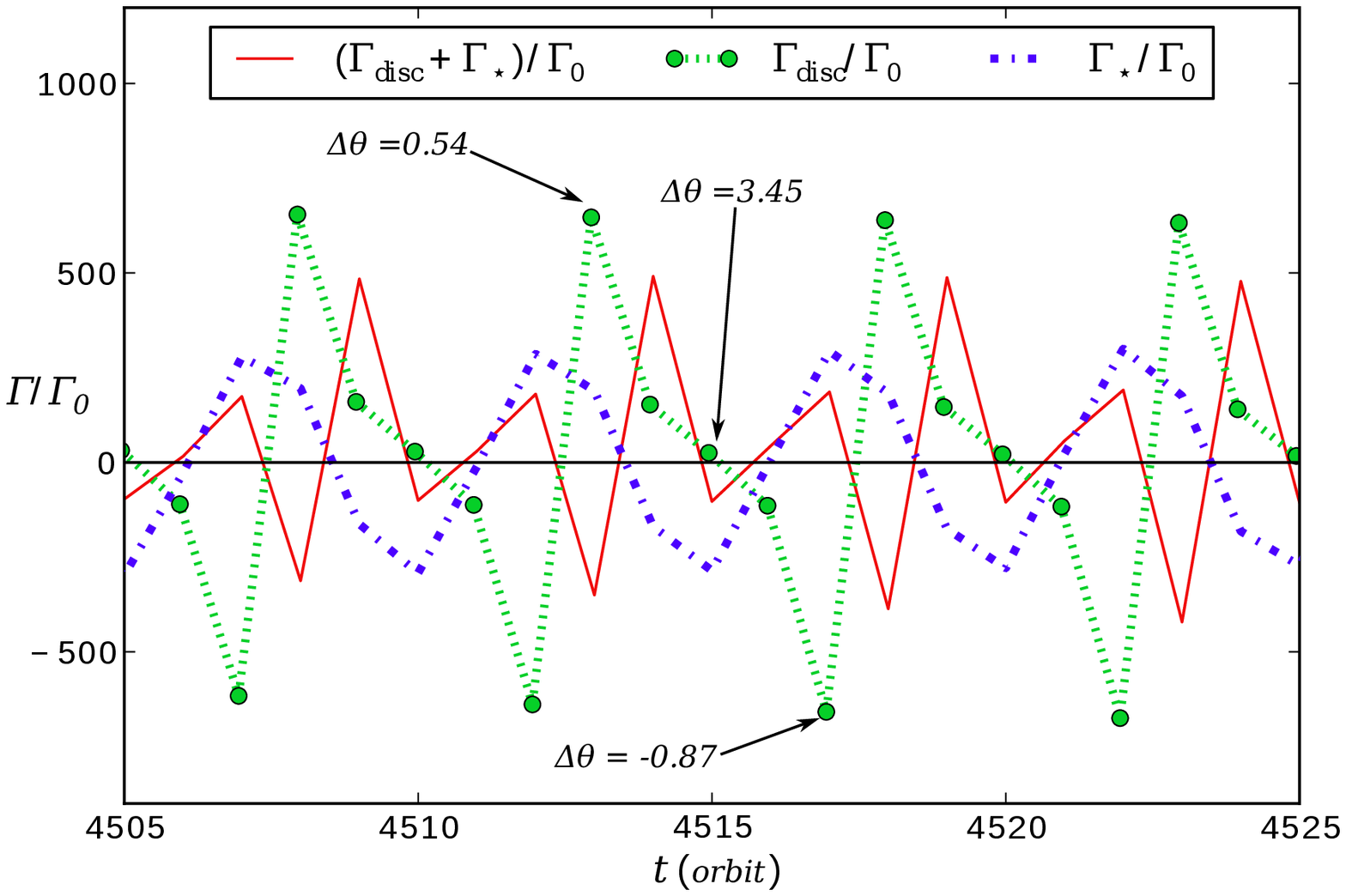}
		\caption{The upper panel displays the scaled torque risen from disc and star on the planet for model Height2. The red and black (thin) solid lines represent the net torque and the time averaged torque respectively. The lower panel shows a shorter section of time between $[4505-4525]$ orbits. $\theta_{v-p}$ in this figure, denotes the azimuthal distance between the vortex and the planet. This value is given for three different times 4512.9, 4517.15 and 4515.15 orbits.}
		\label{h2Torq}
	\end{center}
\end{figure}

To examine whether a more massive vortex is able to trap a planet, we re-ran the standard model with $a=2$ that contained more mass inside the bump. The yellow line in Fig.\ref{h2semi} represents the semi-major axis of the planet against time. In this case, the plant is ''prevented`` from inward migration instead of being trapped. In Fig.\ref{h2Torq}, we display the star, disc and net torques on the planet. After averaging the net torque over the last $4000$ orbits, we plotted this value by the black thin solid line. This figure shows that after the planet inward migration is stopped, the average net torque on the planet is almost zero whereas the net torque at each time does not vanish.

We indicated the azimuthal distance between the planet and the vortex in three different positions in Fig\ref{h2Torq}. Two of them refer to the positions where the disc torque has the maximum and minimum values. Another one is where the disc torque is nearly zero. When the vortex is after the planet ($\theta_{v-p}=0.54$), the disc torque is large and positive. As the vortex rotates and crosses the opposite side of the planet ($\theta_{v-p}=3.45$), the disc torque becomes about zero. And when the vortex comes close to the planet from back, the disc torque changes to a large negative value. The dependency of the sign and the value of the disc torque on the vortex position shows that in this model, the main torque on the planet is due to the gravitational torque of the vortex. Therefore, if the vortex is strong enough, the vortex can play the main role in halting the migration of the planet even if it is not located inside the horseshoe region.

\begin{figure}
	\begin{center}
		\includegraphics[width=9cm]{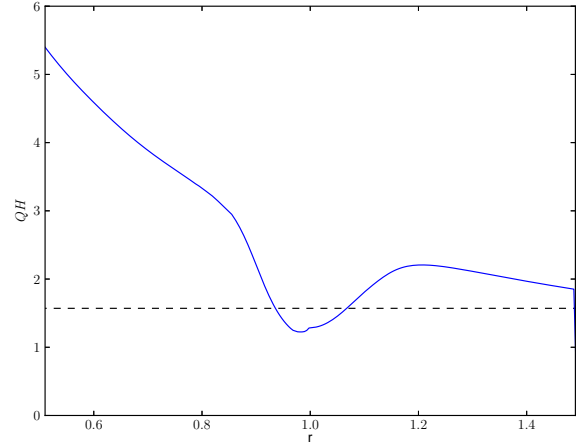}
		\caption{Toomre value multiplied by the disc scale height $QH$ for the Height2 model as a function of radius. $Q$ is calculated at before the vortex formation. The dotted black line shows the critical value $\pi/2$.}
		\label{toomre}
	\end{center}
\end{figure}

Whether self-gravity (SG) alters the results of the Height2 model is an important question. \cite{Lovelace2012} analytically studied the stability condition in thin discs with grooves or bump at the presence of self-gravity. They showed that discs becomes unstable for the axisymmetric perturbation if $Q > (\pi/2) (r_{0}/H)$, or equivalently $Q (H/r_{0})>\pi/2$, where $Q=\kappa c_{s}/\pi G \Sigma$ is Toomre parameter \citep{Toomre1964}, and $\kappa=\left( r^{-3}\partial(r^2\Omega^4)/\partial r \right)^{0.5}$ is the epicyclic frequency. We plotted $QH$ as a function of $r$ before the vortex formation in Fig.~\ref{toomre}. It indicates
that SG might play a role in this model.

\begin{figure}
	\begin{center}
		\includegraphics[width=9cm]{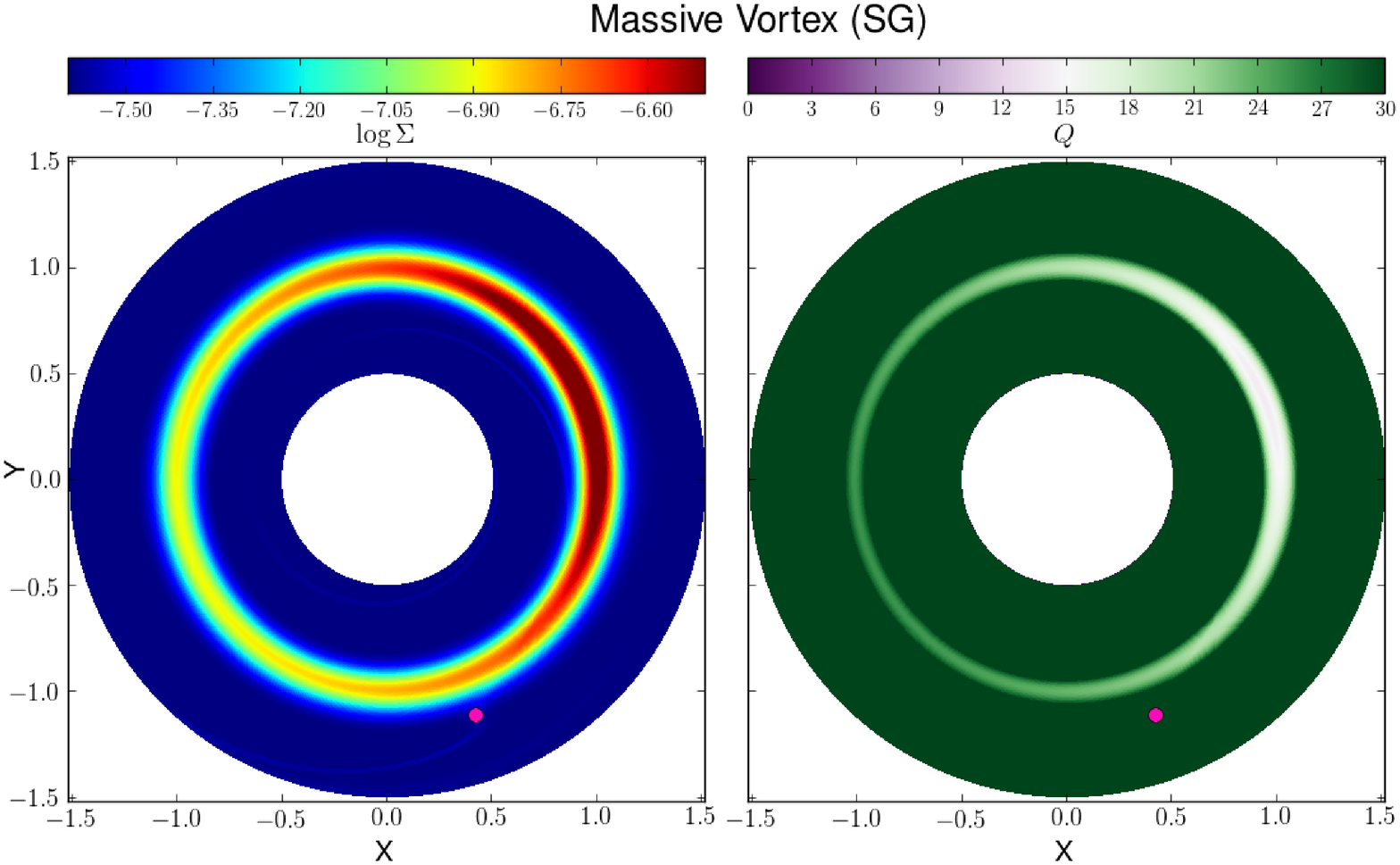}
		\includegraphics[width=9cm]{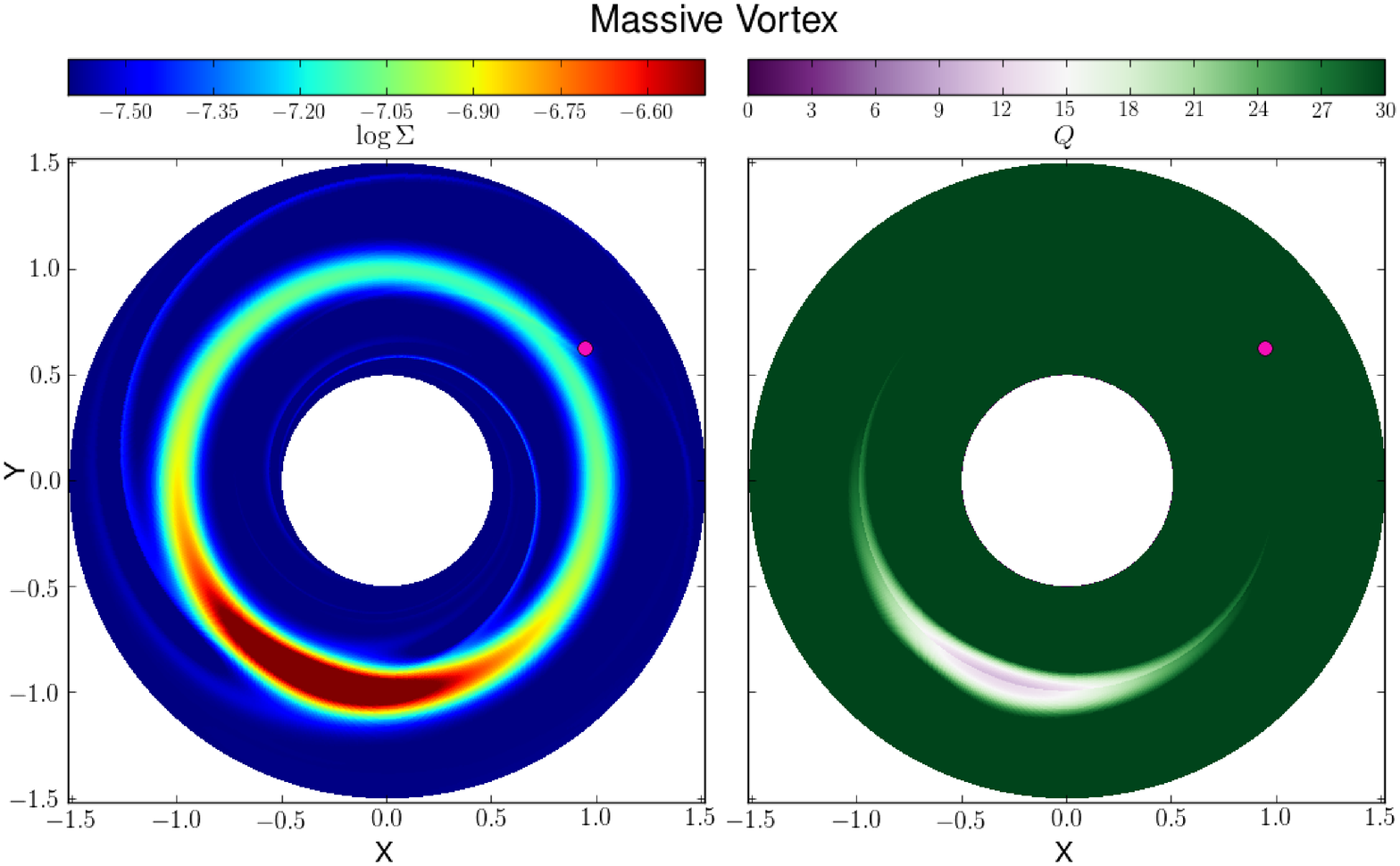}
		\caption{The upper and lower panels respectively present the surface density $\Sigma$ and Toomre parameter $Q$ for the Height2 (top) and Height2~(SG) (bottom) models at $t=13000$~orbits, when the planet migration has been halted.}
		\label{SG2d}
	\end{center}
\end{figure}

To directly test the effect of SG in this case, we repeated an identical simulation with the self-gravitating version of \texttt{FARGO}. 
First we note that the vortex still exits in the SG case. 
With SG the vortex is more elongated but weaker, i.e.
the density at the center of the vortex in this case is lower than in the model without SG.
Moreover, $Q$ is much higher than unity even at the center of vortex (see Fig.~\ref{SG2d}). 
Then we followed the evolution of an embedded planet that started at $a_0\simeq1.2$.
The blue line in Fig.~\ref{h2semi} shows the semi-major axis of the planet for the SG model. Although the migration is slower than the Height2 model because of the disc self-gravity \citep{Pierens2005}, the planet again does not reach the vortex but its inward migration stops due to the gravitational interaction between the vortex, the star and the planet. In Fig.~\ref{SG2d}, we plotted the surface density and Toomre parameter at $t=13000$~orbits for the model with and without SG.

As table~\ref{table:2} suggests, the locking or stopping of the planet depends on the surface density at the vortex center. In the table, we compared the properties of the vortices in different models. The models Visc2 and Dens1 have the most massive vortices because of the very elongated shape in the first model and the higher surface density in the second one. In none of these two models the planet migration is stopped. On the other hand, Height2 and Height2~(SG) models have the highest surface density at the center of their vortices, and in both of them the planet migration is halted farther away from the vortices. This comparison shows the surface density at the vortex center is an important factor in different trapping behaviour of the planet. When the planet comes close to the vortex center, it exchanges torque with the vortex center and if the torque is large enough, it migrates outward as in Height2 models. Because the torque of star and vortex on the planet scale with planet mass while the Linblad torque scales with planet mass squared, a turnover point in the planet's migration behaviour as a function of the planet mass is expected. For planet masses larger than a critical value, the planet could not be expelled by the vortex any more. We leave the detailed analysis of this point to a future study because it is beyond the scope of this paper.

The formation of one massive large scale vortex in the self-gravitated model might appear in contradiction with the result of some other works \citep[e.g.][]{Mamatsashvili2009} that show the vortices sizes are limited by the Jeans scale of the disc. It can be attributed to our specified viscosity profile, which maintains the RWI and vortex formation continuously. \cite{Mamatsashvili2009} argues that the combined effect of self-gravity and Keplerian shear opposes the merging of vortices or destructs the large-scale vortices into smaller ones. However, in \cite{Lyra2009a}, who use a setup more similar to ours both in viscosity reduction and locally isothermal equation of state, large vortices (m=3) with $Q \approx 1$ are formed at $t=75~orbits$. Because in our model, (a) the energy dissipation can not occur because of omitting the energy equation, (b) the RWI is continuously driven by the fixed-shaped bump and the planet, and (c) the simulations are performed for a large number of orbits, the large-scale vortex is formed and survives. 

\begin{figure*}
	\begin{center}
		\includegraphics[width=18cm]{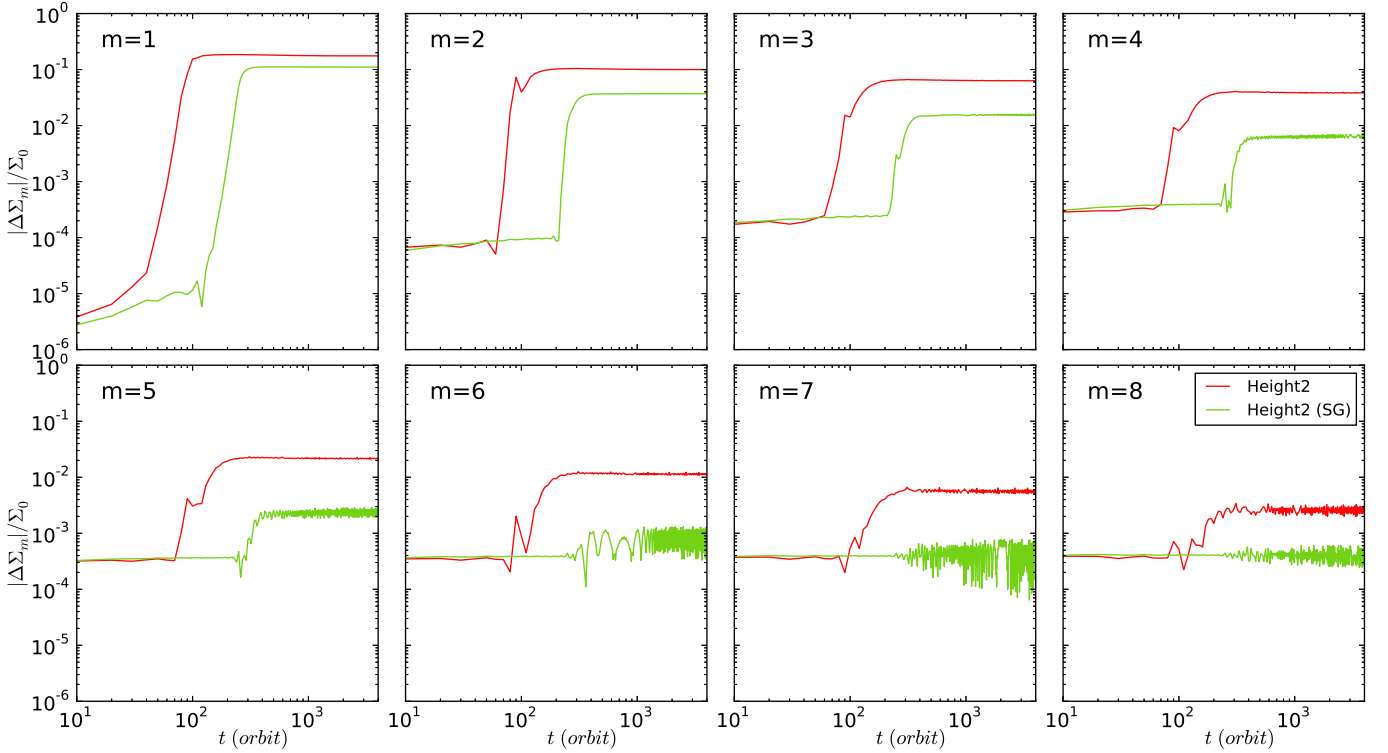}
		\caption{The comparison of different modes between Height2 and Height2~(SG) models.}
		\label{SGnonSGmode}
	\end{center}
\end{figure*}

In Fig.~\ref{SGnonSGmode}, we compared the evolution of different modes of density Fourier transform for self-gravitating and non-self-gravitating models. We expanded the density at the bump tip $\Sigma_{max}(\theta)$ as

\begin{equation}
	\label{fourdens}
	\Sigma_{max}(\theta)=\sum_{m} \Sigma_{m} \exp(-i m \theta)
\end{equation}

\noindent and plotted the difference between $\Sigma_{m}$ and the initial density $\Sigma_{0}$ as a function of time. In both models, the higher modes are dominant initially, but all of them have equal amplitudes. Later, all lower modes grow and finally $m=1$ becomes the strongest. After the linear growth phase, all modes have smaller amplitude in the SG model than the non-SG one. Another important point is that the growth happens in a later time for the SG model. This is in agreement with the results of \cite{Lin2012a} and \cite{Lin2011c} that studies the effect of self-gravity on the vortices formed at the edge of a gap and showed that the self-gravity delays merging of vortices. In our models, although we do not see merging because of mixing the modes, the postponing of vortices growth due to the self-gravity can be observed clearly.
We defer a more detailed study of vortices in SG discs to future work.

\begin{table*}
\caption{Vortex properties for different models. The values are calculated after planet trapping or stopping (T and S in the last column stand for trapping and stopping). $\Sigma_{centre}$, $\Delta \theta$ and $[R_{in},R_{out}]$ represent surface density at the center, azimuthal and radial extension of the vortex.}
\label{table:2}
\centering
\begin{tabular}{c c c c c c}
\hline\hline
Model & $\Sigma_{centre}/\Sigma_{BG}$ & $\Delta \theta$ & $[R_{in},R_{out}]$ \tablefootmark{*}& $M_{vortex}/M_{p}$& T/S\\
\hline
Standard & $2.16$ & $2.8$ & $[0.92,1.05]$ & $32.7$ & T \\
Visc1 & $2.17$ & $3.2$ & $[0.93,1.06]$ & $41.8$ & T\\
Visc2 & $1.97$ & $5.21$ & $[0.92,1.05]$ & $64.7$ & T\\
Dens1 & $2.36$ & $2.12$ & $[0.92,1.05]$ & $49.6$ & T\\
Dens2 & $1.90$ & $1.77$ & $[0.92,1.05]$ & $4.13$ & T\\
Height2 & $4.86$ & $1.85$ & $[0.89,1.09]$ & $36.1$ & S\\
Height2 (SG)& $3.57$ & $2.35$ & $[0.93,1.05]$ & $27.1$& S \\
\hline
\end{tabular}
\end{table*}

\section{Summary and conclusion}
\label{summary}

We studied the interaction between a low mass planet (upto $q=5\times10^{-6}$) and a 2D large-scale stationary vortex formed in a viscously stable Gaussian pressure bump. Whether such bumps do really exist in protoplanetary discs is still an open question whose definite answer needs higher resolution observation. However, the ring structures in transitional discs can be considered as a strong evidence for the presence of pressure bumps and even vortices \citep[e.g.][]{Brown2009,Casassus2013,VanderMarel2013}. Numerical simulations of discs with gap-opening planets show that in a very weakly viscous discs, a density bump as high as $\sim1.8$ times of the background density can be created \citep{Crida2006}. Another probable situation is very narrow viscosity transition in an outer dead-zone edge. In the 2D simulations of \cite{Regaly2013}, the density raised to 5 times of the background value but in a wider distance than the thickness of the disc. If the width of viscosity transition becomes narrower, a bump similar to our models may be created.

Our results show that in all of our basic models the planet migrates toward the vortex, interacts with it and eventually is locked to the vortex in an identical radial distance and at a specific angular position far from the vortex center. The locking position is determined by the balance between the torque from the vortex, star and the local disc. We noticed that if the vortex becomes stronger, the planet exhibits different behaviour. Instead of ''locking``, it is ''stopped`` and its migration is halted farther away from the vortex position.

We also set up the planet inside the vortex in some of our models and we discovered that the planet is expelled out of the vortex during the first hundreds orbits and afterward the planet is locked to one side of the vortex. It can even happen for a planet as low mass as $5\times10^{-7}\mathrm{M_{\star}}$. This has two consequences: the good one is that the vortex can serve as a ''womb`` and continue producing more planetary cores one after another, the dark side is that the planetary core is ejected by the vortex and could not grow to higher mass values.

We caution that in this work we introduced the planet suddenly inside the vortex. \cite{Meheut2012b} show that large dust particles can affect their parent vortices greatly. Thus, a more complete work is needed to answer the question if a vortex can maintain its shape while growing a planet. Another weakness of our work is that we could not study whether a low-mass planet has a role in vortex destruction. This question will be answered in \citeauthor{RegalyPre}(in prep.). 

We performed our models for a fixed pressure bump and consequently a fixed vortex in a flat disc. In more realistic models, the vortex could migrate \citep{Paardekooper2010a}, move in the disc \citep{Regaly2013} or even be threatened by decay \citep[e.g.][]{Meheut2012a}. It is very interesting to explore whether a moving vortex in a more realistic disc is able to "lay a planet system" or not. We will try to answer this question in future works.

\begin{acknowledgements}
We thank \emph{P.~Pinilla} for the fruitful discussions, \emph{F.~Masset} for his kind help in interpreting \texttt{GFARGO} output files and \emph{J.~Ramsey} for his kind help in maintaining the GPUs in Institute for Theoretical Astrophysics. This work was financed partly by a scholarship of the Ministry of Science, Research, and Technology of Iran, and partly by the 3rd funding line of the German Excellence Initiative. Zs.~Reg\'{a}ly has been supported by the Lend\"{u}let-2009 Young Researcher's Program of the Hungarian Academy of Sciences, City of Szombathely under agreement No.K-111027 and Hungarian grant K-101393.
\end{acknowledgements}
\bibliographystyle{aa} 
\bibliography{lib.bib} 

\appendix 
\section{Torques at the vortex center}
\label{app1}
On a planet inside a vortex, only Lindblad, vortex gravitational and star torques can play role. The corotation torque needs the horseshoe orbits which can not be formed inside the vortex due to the circulation of material inside the vortex. Therefore, this component can be ignored until the planet comes out of the vortex. 

\begin{figure}
	\begin{center}
		\includegraphics[width=9cm]{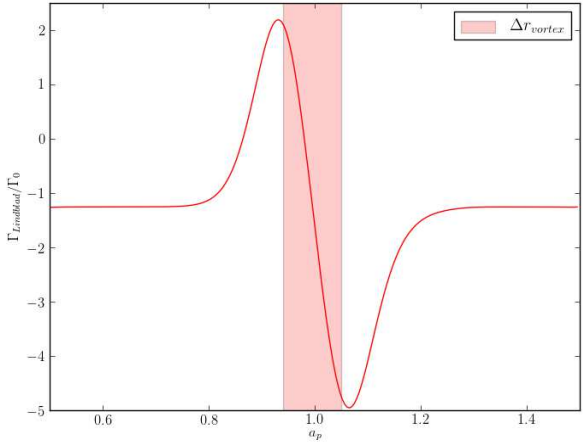}
		\caption{The maximum Lindblad torque at planet orbital position. The torque is calculated using relation~\ref{lindmax} and the radial density profile of the disc before vortex formation. The Lindblad torque has the highest value when planet is close to the bump edges. The pink area shows the vortex radial extension in the bump after it formed. These values are calculated for the Visc1 model.}
		\label{LindAna}
	\end{center}
\end{figure}

The Lindblad torque is calculated by summing the individual torques from each Lindblad resonances. \cite{Masset2011} showed that the Lindblad torque is not very sensitive to the second derivative of surface density. Therefore, we used the analytical formula by \cite{Paardekooper2010} to estimate the maximum Lindblad torque in every position of the disc with a symmetric bump. Figure \ref{LindAna} shows the the Lindblad torque on the planet calculated by

\begin{equation}
	\label{lindmax}
	\frac{\Gamma_{Lindblad}}{\Gamma_{0}}= -(2.5+1.7\beta-0.1\alpha)(\frac{0.4}{\epsilon})^{0.71}
\end{equation}

\noindent{where $\alpha=-\mathrm{d}\log{\Sigma}/\mathrm{d}\log{r}$ and $\beta=-\mathrm{d}\log{T}/\mathrm{d}\log{r}$ which equals to $1$ in our models. We calculated $\alpha$ numerically at each $r$ over the azimuthally averaged surface density. The figure shows that the scaled Lindblad torque equals to $-2$ for a planet at the bump center. Based on the definition of Lindblad torques that uses the epicylic frequency, in our models, the Lindblad resonances can be formed outside of the vortex and consequently farther away from the planet. This can even lower the normalized Lindblad torque $\Gamma/\Gamma_{0}$ to a value less than $-2$.}

\begin{figure}
	\begin{center}
		\includegraphics[width=9cm]{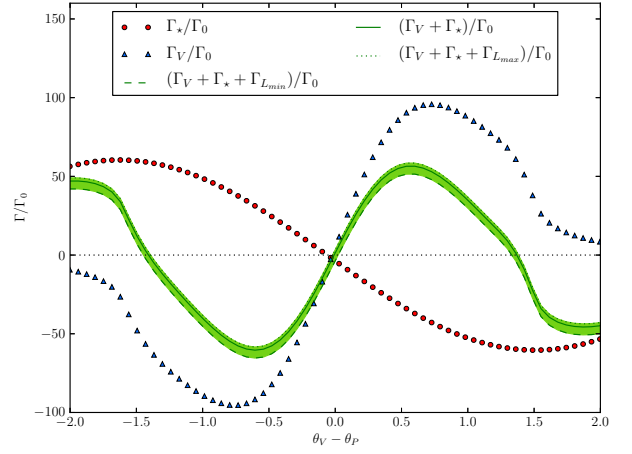}
		\caption{The normalized star and vortex gravitational torques on the planet at different azimuthal position toward the vortex center at the bump maxima. The green solid line represents the sum of the two components. The dashed and  dotted lines demonstrate the variation of the total torque by the Lindblad torque. The maximum and minimum values of the Lindblad torque are obtained from Fig.~\ref{LindAna}. The horizontal axis is limited to the vortex azimuthal extension. }
		\label{Caltorq}
	\end{center}
\end{figure}

To calculate the star and vortex gravitational torques, we put the planet at different azimuth with respect to the vortex center at the bump maximum and calculated these two components for Visc6 model. Noting that the Lindblad torque is much smaller than the vortex gravitational and star torques, it can only shift the zero torque position to the right (trailing) or left (leading) of vortex center. As Fig.\ref{Caltorq} shows, if the planet is displaced slightly to the right of the vortex center, the positive torque causes outward migration and if the planets moves slightly to the left, the negative torques bring the planet out of vortex. Except the Lindblad torques, which has minor effect in our models (see the green area in Fig.~\ref{Caltorq}), all the other normalized torque components are independent of planet mass. What is the lowest mass planet that can remain inside the vortex is a question that can be answered by studying the planet formation from dust to planet that is not the scope of this paper.

\end{document}